\documentclass{emulateapj}
\usepackage{graphicx}

\begin{document}

\def\etal{et al.\ \rm}
\def\ba{\begin{eqnarray}}
\def\ea{\end{eqnarray}}
\def\etal{et al.\ \rm}
\def\Fdw{F_{\rm dw}}
\def\Tex{T_{\rm ex}}
\def\Fdis{F_{\rm dw,dis}}
\def\Fnu{F_\nu}
\def\FJ{F_{J}}
\def\FJE{F_{J,{\rm Edd}}}

\title{Structure and evolution of circumbinary disks around 
supermassive black hole (SMBH) binaries.}

\author{Roman R. Rafikov\altaffilmark{1,2}}
\altaffiltext{1}{Department of Astrophysical Sciences, 
Princeton University, Ivy Lane, Princeton, NJ 08540; 
rrr@astro.princeton.edu}
\altaffiltext{2}{Sloan Fellow}


\begin{abstract}
It is generally believed that gaseous disks around 
supermassive black hole (SMBH) binaries in centers of galaxies 
can facilitate binary merger and give rise to  
observational signatures both in electromagnetic and 
gravitational wave domains. We explore general properties of 
circumbinary disks by reformulating 
standard equations for the viscous disk evolution in terms
of the viscous angular momentum flux $F_J$. In steady state 
$F_J$ is a linear function of the specific angular 
momentum, which is a generalization of (but is not 
equivalent to) the standard constant $\dot M$ disk solution. 
If the torque produced by the central binary is effective at 
stopping gas inflow and opening a gap (or cavity) in the disk, 
then the inner part of the circumbinary disk can be approximated 
as a constant $F_J$ disk. We compute properties of such disks in 
different physical regimes relevant for SMBH binaries and use 
these results to understand the gas-assisted evolution of SMBH 
pairs starting at separations $10^{-4}-10^{-2}$ pc. We find the 
following. (1) Pile-up of matter at the inner edge of the disk 
leads to continuous increase of the torque acting on the binary and
can considerably accelerate its orbital evolution compared to 
the gravitational wave-driven decay. (2) Torque on the binary 
is determined non-locally and does not in general
reflect the disk properties in the vicinity of the binary. 
(3) Binary evolution depends on the past history of the disk 
evolution. (4) Eddington limit can be important in circumbinary 
disks at late stages of binary evolution even if they accrete 
at sub-Eddington rates. (5) Circumbinary disk self-consistently 
evolved under the action of the binary torque 
emits more power and has spectrum different from the 
spectrum of a constant $\dot M$ disk 
--- it is steeper ($\nu F_\nu\propto \nu^{12/7}$) and 
extends to shorter wavelength, facilitating its detection. 
Our results can be used for 
understanding properties of circumbinary disks in other 
astrophysical settings.
\end{abstract}



\section{Introduction.}  
\label{sect:intro}

Physics of gaseous disks around astronomical objects, also 
known as accretion disks, has been one of the most important 
topics in astrophysics since the pioneering works of 
Shakura \& Sunyaev (1973), Novikov \& Thorne (1973), 
Lynden-Bell \& Pringle (1974; hereafter LBP74). Historically,
a lot of attention has been paid to understanding the 
properties of disks in which the mass accretion rate
$\dot M$ is constant with radius. Some other varieties of 
disks have been studied as well, in particular the so-called 
{\it circumbinary disks} 
--- gaseous disks orbiting a central binary, which can be 
a stellar binary or a pair of supermassive black holes (SMBHs)
in centers of galaxies. The latter type of systems has recently 
attracted a lot of attention since the tidal interaction 
of the SMBH binary with the disk results in the loss of the  
orbital angular momentum of the former and its faster inspiral. 
This may help resolve the so-called ``last parsec'' 
problem (Yu 2002; Lodato \etal 2009) --- stalling of the 
SMBH binaries at separations of $10^{-3}-1$ pc caused by 
the inefficiency of stellar dynamical processes at shrinking 
their orbits. Such circumbinary 
disks are the prime focus of this work.

Depending on the mass ratio of the binary components one
can have different modes of tidal coupling of the binary 
with the disk. When the secondary-to-primary mass ratio 
$q\equiv M_s/M_p$ is very small, the secondary cannot perturb 
the disk significantly and migrates through it in the 
so-called Type I migration regime familiar from the studies 
of protoplanetary disk-planet interaction (Ward 1997). 
At higher mass ratios\footnote{The exact transition between the 
Type I and II regimes depends not only on $q$ but also on the 
viscosity and other disk parameters, see Lin \& Papaloizou (1986),
Rafikov (2002).} the secondary becomes capable of clearing 
gas from the annulus around its orbit, switching its orbital
evolution into the so-called Type II migration regime.
As $q$ gets closer to unity the gap around the orbit
of the secondary turns into a central cavity inside of which 
the binary resides. Numerical simulations (MacFadyen \& 
Milosavljevi\'c 2008; Cuadra \etal 2009) typically show 
that for $q\sim 1$ the size of the cavity is about twice the
semi-major axis of the binary orbit.

Deposition of the angular momentum of the density waves 
excited by the binary causes non-trivial evolution of the 
circumbinary disk. The seminal study of LBP74 followed 
by works of e.g. Lightman \& Eardley (1974), 
Lin \& Papaloizou (1996) have shown that 
the behavior of disks evolving under the action of both 
the internal viscosity and the external torque can 
be described by a simple equation
\ba
\frac{\partial \Sigma}{\partial t}=-\frac{1}{r}
\frac{\partial}{\partial r}\left[
\left(\frac{\partial l}
{\partial r}\right)^{-1}\frac{\partial}{\partial r}
\left(r^3\nu\Sigma\frac{\partial\Omega}
{\partial r}\right)+\frac{2\Sigma\Lambda}{\Omega}\right],
\label{eq:evSigma}
\ea
where $\Lambda$ is the external torque per unit mass
of the disk. Here $\Sigma$ and $\nu$ are the surface density 
and kinematic viscosity of the disk, $t$ and $r$ 
are time and radius, and $l=\Omega(r)r^2$ is 
the specific angular momentum for circular orbit. 

In this work, following  LBP74 and Lyubarskij 
\& Shakura (1987) we reformulate equation (\ref{eq:evSigma}) in a 
more convenient form which allows a very straightforward 
interpretation of the steady state disk structure and
provides a transparent way of understanding the evolution
of circumbinary disks. This allows us to describe the 
structure and electromagnetic signatures of the circumbinary 
disks around SMBH binaries, their variation in time, and 
the back-reaction onto the orbital evolution of the central 
binary, leading to its inspiral.

Our work is structured as follows. In \S \ref{sect:evolution}
we provide a general description of the coupled disk-binary 
evolution, derive governing equations, and obtain steady state  
solutions that generalize previously known results. In \S 
\ref{sect:SMBH} we provide the description of the properties 
of disks around SMBH binaries parametrizing them via the 
viscous angular momentum flux rather than the mass accretion
rate $\dot M$. This parametrization allows simple
description of the properties of steady state circumbinary 
disks. In \S \ref{sect:evol} we discuss evolution of 
circumbinary disks, in particular deriving the self-similar 
solutions to an evolving disk structure. All these results 
are then used in \S \ref{sect:SMBHs} to describe the 
coupled evolution of a SMBH binary and a circumbinary disk,
including the self-consistent time variation of the disk 
properties, the orbital inspiral of the binary components,
and the electromagnetic manifestations of the system.  
Our main results are summarized in \S \ref{sect:summ}.


\section{Problem setup and evolution equation.}  
\label{sect:evolution}

We consider a binary consisting of two point masses $M_p$
and $M_s<M_p$ (total mass is $M_c=M_p+M_s$), moving around 
common barycenter on circular orbits (for simplicity we neglect the 
possibility of non-zero eccentricity). Binary is surrounded by 
a proghrade, coplanar disk (cf. Nixon \etal 2011) which 
extends to much larger distances than the binary 
semi-major axis $r_b$. We assume the eccentricity of the binary
to be negligible, even though simulations show the possibility
of eccentricity growth due to the tidal binary-disk coupling
(Roedig \etal 2011).

The disk is truncated by the binary
torque at the radius $r_{in}$; the width of the 
gas-depleted annulus between the orbit of the secondary and 
$r_{in}$ is $\Delta\equiv|r_{in}-r_b|\lesssim r_b$ for $q\ll 1$ 
while $\Delta\sim r_b$ for $q\sim 1$ (MacFadyen \& 
Milosavljevi\'c 2008). For simplicity we assume 
that neither the primary nor the secondary have their own 
disks (this simplification can be easily relaxed). The 
detailed conditions for gap opening by a massive perturber 
in a disk can be found elsewhere (Lin \& Papaloizou 1986,
Rafikov 2002).  Here we just assume that the binary 
torque clears out a clean central 
cavity in the disk. In practice this generally sets a lower 
limit on the mass ratio of the binary components $q$ (determined
by the local conditions in the disk), below which the tidal 
torque of the secondary is too weak to prevent viscous refilling 
of the gap (or cavity). We determine the conditions under
which such an overflow of the disk across the orbit of 
the secondary (Kocsis \etal 2012a,b) is possible in 
\S \ref{sect:overflow}.

External torque $\Lambda$ due to tides raised by the central 
binary is expected to be concentrated right at the edge 
of the gap or cavity around the orbit of the secondary. 
This expectation is borne out in calculations of
the torque density distribution both in uniform disks,
where it is generally found that $\Lambda\propto |r-r_b|^{-4}$
(Goldreich \& Tremaine 1980; Armitage \& Natarajan 2002),
and in nonuniform disks, where it has been shown by 
Petrovich \& Rafikov (2012) that $\Lambda$ decays 
{\it exponentially} near the disk edge. As a result, already at 
small separations from $r_{in}$ the term 
proportional to $\Lambda$ in equation 
(\ref{eq:evSigma}) can be neglected; this equation then 
reduces to its classical form first obtained by LBP74. 
Effect of the binary torque is then incorporated in the solution 
of this simplified equation via the boundary condition 
(discussed in \S \ref{sect:BCs}) imposed at the inner 
edge of the disk.

Throughout this work we will assume the potential in which the
disk orbits to be faithfully represented by a Newtonian potential
produced by a combined mass $M_c$. 

Evolution of an accretion disk is best illustrated if we 
characterize the disk at each radius not by its surface density
$\Sigma$ but by the viscous angular momentum flux $\FJ$ 
defined as
\ba
\FJ\equiv -2\pi\nu\Sigma r^3\frac{d\Omega}{dr}=3\pi\nu\Sigma
\Omega r^2, 
\label{eq:Fnu}
\ea
where the last equality is for Keplerian disks with 
$\Omega=(GM_c/r^3)^{1/2}$.
This quantity represents the viscous torque exerted by the 
part of the disk interior to a given radius $r$ on the external part of 
the disk, and is thus equal to the amount of angular momentum 
crossing the disk circumference $2\pi r$ per unit time due
to the action of viscosity.

In the absence of external torques mass accretion rate 
through the disk can be directly expressed (LBP74)
through the divergence of the viscous angular momentum flux 
as (note that we take $\dot M>0$ for mass inflow towards the 
center of the system)
\ba
\dot M(r)=\left(\frac{dl}{dr}\right)^{-1}
\frac{\partial \FJ}{\partial r}=\frac{\partial \FJ}{\partial l},
\label{eq:Mdot}
\ea 
motivating us to change the independent variable from $r$ to $l$. 
In this case the evolution equation (\ref{eq:evSigma}) with 
$\Lambda=0$ takes on a particularly simple form (LBP74; Filipov 1984; 
Lyubarskij \& Shakura 1987):
\ba
\frac{\partial }{\partial t}\left(\frac{\FJ}{D_{J}}\right)=
\frac{\partial^2 \FJ}{\partial l^2},
\label{eq:evF}
\ea
where the function similar to the diffusion coefficient 
\ba
D_{J}\equiv -\nu r^2\frac{d\Omega}{dr}\frac{dl}{dr}
\label{eq:D_J}
\ea
is in general a function of both $l$ and $\FJ$ because of
the possible dependence of $\nu$ on $\FJ$. 

Despite the mathematical simplicity of the evolution equation 
(\ref{eq:evF}), which was first recognized by LBP74, over the 
years it has become conventional to study disk evolution using
the more complicated equation (\ref{eq:evSigma}). We show next 
that use of equation (\ref{eq:evF}) allows certain advantages 
over the standard approach, in particular for obtaining the 
steady state solutions for the disk structure.


\subsection{Boundary conditions.}  
\label{sect:BCs}

Using equation (\ref{eq:evSigma}) and continuity equation 
one finds
\ba
-\dot M=-\left(\frac{\partial l}
{\partial r}\right)^{-1}\frac{\partial \FJ}{\partial r}
+4\pi\frac{\Sigma\Lambda}{\Omega}.
\label{eq:dotM_gen}
\ea
As we mentioned before, $\Lambda$ is expected to be
significant only in a narrow annulus at the inner edge of
the disk. We may then assume for simplicity that $\Lambda=0$
outside of some radius $r_\Lambda$, which is not too different 
from $r_{in}$ (to be specific, one can e.g. take $r_\Lambda$ 
to be the radius interior to which the binary exerts $90\%$ 
of its torque on the disk). Multiplying equation 
(\ref{eq:dotM_gen}) by $dl/dr$ and integrating between 
$r_b$ and $r_\Lambda$ one gets
\ba
\FJ(r_\Lambda)=2\pi\int\limits_{r_b}^{r_\Lambda}
r\Lambda\Sigma dr+\frac{1}{2}
\int\limits_{r_b}^{r_\Lambda}
\dot M\Omega r dr,
\label{eq:intMdot}
\ea
where we set $\FJ(r_b)=0$ because $\Sigma(r_b)=0$ by our 
assumption of a clean gap.

The first integral on the right hand side of this expression
is the total torque that the binary exerts on the disk. 
As long as the orbital evolution of the binary is driven 
predominantly by the tidal coupling to the disk (and not due to the
gravitational wave emission) conservation of the angular 
momentum ensures that this term is equal to 
\ba
-\frac{dL_b}{dt}=-\frac{L_b}{2r_b}v_b, 
\label{eq:v_b}
\ea
where $L_b=M_c(GM_c r_b)^{1/2}q/(1+q)^2$ is the orbital angular 
momentum of the binary and $v_b\equiv dr_b/dt$ is its inspiral
speed. 

In the second integral in equation (\ref{eq:intMdot}) one can 
write $\dot M=2\pi\Sigma r v_r$,
where $v_r$ is the radial velocity of the gas, and approximate
$v_r\sim v_b$ in the annulus between $r_b$ and $r_\Lambda$, thus
assuming that gas in this annulus closely follows the shrinkage 
of the binary orbit. Then one can easily see that as long as the 
mass of the secondary $M_s$ satisfies the condition 
\ba
M_s\gtrsim \frac{1}{\Omega(r_b) r_b}\int\limits_{r_b}^{r_\Lambda}
\Sigma\Omega r^2 dr,
\label{eq:M_s_limit}
\ea
the second term in the right hand side of equation 
(\ref{eq:intMdot}) is smaller than the first one. 
This condition is often replaced by demanding that the 
``local disk mass'' $M_d\equiv \Sigma r^2$ in the vicinity of the 
binary (i.e. at $r\sim r_b$) be less than the mass of
the secondary (SC95, hereafter SC95; Haiman \etal 2009). 

Whenever we are in the limit (\ref{eq:M_s_limit}) the viscous
angular momentum flux in the inner region of the disk can
be directly related to the orbital evolution of the binary:
\ba
\FJ(r_{in})=-\frac{dL_b}{dt}.
\label{eq:cond}
\ea
Here we replaced $r_\Lambda$ with $r_{in}$ since the two 
radii are very similar (and also very close to $r_b$). 
On the contrary, if the condition 
(\ref{eq:M_s_limit}) is not fulfilled the inner part of the 
disk absorbs most of the angular momentum brought in by viscous
torques and equation (\ref{eq:cond}) becomes invalid. In this
case a more general boundary condition in the form 
(\ref{eq:intMdot}) must be employed. 

Using definition (\ref{eq:Fnu}) and expression for $L_b$ 
one can rewrite equation (\ref{eq:cond}) as a formula for
the orbital evolution time of the binary $t_{ev}=|d\ln r_b/dt|^{-1}$:
\ba
t_{ev}=\frac{L_b}{2\FJ(r_{in})}=\frac{t_\nu}{6\pi(1+q)}
\frac{M_s}{M_d},
\label{eq:t_ev}
\ea
where we defined the viscous timescale $t_\nu\equiv r^2/\nu$, 
evaluated at $r_b$. Similar expressions (up to a constant factor) 
have previously been quoted in the literature on
Type II migration in the limit of $M_d\ll M_s$, see e.g. 
Lodato \etal (2009), Baruteau \& Masset (2012). 
A subtle point in this expression is that both $t_\nu$
and $M_d$ vary in time as the disk evolves even if the 
orbit of the secondary does not change appreciably. Calculation 
of this evolution is one of the goals of our present work.

Previously SC95 have used equation (\ref{eq:cond})
coupled with the condition $v_r\to v_b$ as $r\to r_{in}$
as a boundary condition for the problem of the circumbinary disk 
evolution. They have also assumed that a steady state
solution for the disk structure can always be obtained even 
far from the binary, despite the fact that the radius $r_b$ 
(at which the inner boundary condition is imposed)
evolves as the secondary migrates. Because of this inconsistency
their solution explicitly depends on time via the dependence 
on $r_b(t)$ (and the torque on the binary is found to depend on 
the mass of the secondary) and should evolve on the migration 
timescale of the secondary. However, the latter is much shorter 
than the viscous timescale of the disk far from the binary, 
where this solution is still assumed to be valid (by construction, 
the migration timescale of the secondary is of order the viscous 
timescale at $r_{in}\approx r_b$). As a result, 
viscous transport in the disk is unable to communicate 
information about the changing inner boundary condition outside 
the immediate vicinity of the inner edge of the disk, 
implying internal inconsistency of the quasi-steady solution 
derived in SC95. This issue has been first noted by Ivanov \etal 
(1999; hereafter IPP); nevertheless the SC95 solution is still 
being used in studies of the disk-assisted SMBH binary evolution
(Haiman \etal 2009; Kocsis \etal 2011; Yunes \etal 2011).

In our work we take a different approach and specify the 
inner boundary condition in the form of the constraint 
on the mass accretion rate $\dot M$. Previously, IPP 
have used boundary conditions similar 
to ours, even though they have formulated them via the 
asymptotic behavior of the disk surface density 
$\Sigma$. For that reason many of our results coincide with 
their findings. 

At large separations we normally take the disk to be a standard 
constant $\dot M$ disk with the mass supply rate $\dot M_\infty$. 
But close to the binary, as $r\to r_{in}$, $\dot M$
does not in general have to be equal to $\dot M_\infty$.
If the tidal interaction with the binary 
presents a strong barrier for the gas inflow, then 
$\dot M(r_{in})$ is vanishingly small (Liu \& Shapiro 2010). 
But in general tidal torques do not have to completely stop the 
mass inflow: in some situations (e.g. if the gap cleared
out by the secondary is not deep/broad enough to present a 
serious obstacle to the gas inflow) a fraction of mass arriving at 
the inner edge of the disk can cross the orbit of the secondary 
and be accreted by one of the binary components 
(Kocsis \etal 2012a,b). Alternatively, gas can be removed 
from the circumbinary disk in the form of a wind, see 
\S \ref{sect:F_J}, \ref{sect:Edd_lim}. To account for this 
possibility we generally use an inner boundary condition in the 
form 
\ba
\frac{\partial\FJ}{\partial l}\Big|_{r=r_{in}}=\dot M(r_{in})=
\chi\dot M_\infty,
\label{eq:inner_BC0}
\ea
see equation (\ref{eq:Mdot}). Here $\chi\le 1$ is assumed to be 
constant, allowing $\dot M(l_{in})$ to be less than 
$\dot M_\infty$. In practice the value of $\dot M(r_{in})$ and 
$\chi$ is set by the strength of the tidal barrier 
(Liu \& Shapiro 2010) and may vary in time as the system 
evolves. 

Clearly, $\chi=0$ implies no gas inflow across the orbit of 
the secondary, and this is the situation that we will often
consider in this work. For $\chi=1$ secondary does not present any 
barrier to the mass inflow and the disk structure  
reduces to that of a disk with constant $\dot M=\dot M_\infty$.


\subsection{Steady state solution.}  
\label{sect:steady}

Equation (\ref{eq:evF}) clearly admits a simple steady 
state solution (LBP74)
\ba
\FJ(l)=F_{J,0}+F_{J,1} l,
\label{eq:Fsol}
\ea
where $F_{J,0}$ and $F_{J,1}$ are constants, and $l$ is the specific 
angular momentum. This solution is 
completely independent of the detailed physics that determines
the disk properties, since for any (even highly nonlinear) 
dependence of $D_J$ on $\FJ$ and $l$ the solution (\ref{eq:Fsol})
still satisfies equation (\ref{eq:evF}) in steady state.

According to equation 
(\ref{eq:Mdot}) this solution implies
$\dot M=F_{J,1}$ also being constant, which seems to suggest that 
equation (\ref{eq:Fsol}) corresponds to the conventional 
accretion disk with constant $\dot M$. This however is 
not true is general.

First description of a constant $\dot M$ accretion disk 
was provided in a seminal work of Shakura \& Sunyaev (1973), 
who explored properties of disks affected only by the internal 
viscous stresses all the way to the central object. 
They have shown in particular that in such disks the local 
surface density $\Sigma(r)$ at each radius is related to 
the mass accretion rate via
\ba
\dot M=3\pi\nu\Sigma,
\label{eq:Mdot_SS}
\ea 
while the energy loss per unit surface area of the disk scales as 
\ba
\sigma T_{\rm eff}^4(r)=\frac{3}{8\pi}\dot M\Omega^2.
\label{eq:T_SS}
\ea

Over the years the concept of constant $\dot M$ disks has
evolved to essentially imply disk properties given by 
equations (\ref{eq:Mdot_SS}) and (\ref{eq:T_SS}). 
In the rest of this work we will call such disks  
{\it standard constant  $\dot M$} disks.

Let us now  
consider a steady state solution in the form 
(\ref{eq:Fsol}) with $F_{J,0}\neq 0$ and $F_{J,1}=\dot M=const$. 
Combining equations (\ref{eq:Fnu}) and (\ref{eq:Fsol})
one finds 
\ba
\dot M=3\pi\nu\Sigma-F_{J,0} l^{-1},
\label{eq:Mdot_gen}
\ea 
which reduces to (\ref{eq:Mdot_SS}) only when $F_{J,0}=0$
and $\FJ=\dot M l$,
and results in a quite different expression for $\dot M$ otherwise. 
In particular, it follows directly from (\ref{eq:Mdot}) that 
it is possible to have a steady disk with $\dot M=0$ as 
long as $\FJ$ is independent of radius. In such constant $\FJ$ 
disk surface density is related to $\FJ$ via
\ba
\FJ=F_{J,0}=3\pi\nu\Sigma(r)l,
\label{eq:dotM0}
\ea
which replaces equation (\ref{eq:Mdot_SS}).
We explore properties of such disks in
\S \ref{sect:properties}.

Similarly, the viscous energy dissipation rate in the disk 
per unit radius (and per unit time) $d\dot E/dr$ is given by 
\ba
\frac{d\dot E}{dr}=-\FJ\frac{d\Omega}{dr}=\frac{3}{2}
\frac{\FJ\Omega}{r}.
\label{eq:nrg}
\ea
Since $d\dot E/dr=4\pi r\sigma T_{\rm eff}^4(r)$ we can write
\ba
\sigma T_{\rm eff}^4(r)=\frac{3}{8\pi}\frac{\FJ\Omega}{r^2},
\label{eq:T_gen}
\ea
which in steady state described by the solution (\ref{eq:Fsol})
yields
\ba
\sigma T_{\rm eff}^4(r)=\frac{3}{8\pi}\left[\dot M\Omega^2(r)+
\frac{F_{J,0}\Omega}{r^2}\right].
\label{eq:T_steady}
\ea
Again, this expression reduces to the conventional result 
(\ref{eq:T_SS}) only when $F_{J,0}=0$. In 
the case of $\dot M=0$ disk with $\FJ=F_{J,0}$ one finds
\ba
T_{\rm eff}(r)=\left(\frac{3}{8\pi}
\frac{F_J\sqrt{GM_c}}{\sigma}\right)^{1/4}
r^{-7/8},
\label{eq:Tscale}
\ea
so that $T_{\rm eff}(r)$ increases towards small radii more 
steeply than in a standard constant $\dot M$ disk (for which
$T_{\rm eff}(r)\propto r^{-3/4}$). This result 
was first obtained by SC95.

Equation (\ref{eq:T_gen}) predicts non-zero $T_{\rm eff}$
for a disk with $\FJ=const$, even though $\dot M=0$ in such a disk.
Since there is no inward mass flow in this disk and 
corresponding release of potential energy is absent one 
may naturally wonder where does the energy emitted from the
disk surface come from. Integrating equation
(\ref{eq:nrg}) between arbitrarily chosen inner and outer 
radii $r_i$ and $r_o$ one finds the global energy release
between these radii 
\ba
\dot E(r_i<r<r_o)
=\FJ\left[\Omega(r_i)-\Omega(r_o)\right],
\label{eq:dotEnu}
\ea
where we have used the fact that $F_J$ is independent of $r$.
Thus, the global rate of energy generation by viscous
dissipation is equal to the work done on the disk by the 
viscous stress at its inner and outer edges. The latter ultimately 
provides the energy source for the radiation from the disk surface.

To summarize, the assumption of a steady state does not
necessarily require $\dot M$ and $T_{\rm eff}$ in a 
disk to be given by standard equations (\ref{eq:Mdot_SS}) 
and (\ref{eq:T_SS}) --- these equations are sufficient but 
not necessary characteristics of steady disks. The most general
steady state is in fact described by equation (\ref{eq:Fsol}) 
and then it follows that $\dot M$ and $T_{\rm eff}$
must be given by equations (\ref{eq:Mdot_gen}) and 
(\ref{eq:T_steady}). In the rest of the paper we explore 
the properties of such ``non-standard'' disks.


\section{Disk properties as a function of $\FJ$.}  
\label{sect:properties}

It is conventional to describe the structure of steady state 
accretion disks using the mass accretion rate $\dot M$, assumed 
to be constant with radius, as a free parameter (e.g. Shakura 
\& Sunyaev 1973). Such
calculations universally assume that relations 
(\ref{eq:Mdot_SS}) and (\ref{eq:T_SS}) hold true, i.e. that 
$F_{J,0}=0$ in equation (\ref{eq:Fsol}). Having shown in \S 
\ref{sect:steady} that in general steady state does not 
require $F_{J,0}=0$ we now revise these scalings by expressing
disk properties in terms of $\FJ$, which may in general be a 
function of both time and radius $r$ or specific angular 
momentum $l$, rather than $\dot M$. When an additional 
assumption $\FJ=const$ is made these relations naturally 
describe the properties of $\dot M=0$ disks, see \S 
\ref{sect:steady}.

In our calculations we will always make an assumption of an
optically thick disk, but it is easy to extend them
to optically thin disks as well. Because of the large
range of radii spanned by disks in some systems, the
conditions in them are expected to vary with radius, resulting
in transitions in the opacity behavior. To account for this 
in Appendix \ref{sect:gen_opacity} we derive a set of scaling 
relations applicable to the gas pressure dominated disks with
power law opacity behavior, and apply them in \S 
\ref{sect:SMBH} to disks around SMBH binaries.


\subsection{Circumbinary disks around SMBH binaries.}  
\label{sect:SMBH}

Structure of the inner region of a circumbinary disk around a 
SMBH binary may be strongly affected by the radiation pressure. 
We parametrize its role by a dimensionless ratio of the gas 
pressure to total pressure $\beta\equiv p_g/p=p_g/(p_r+p_g)$, where 
$p_g=\rho kT/\mu$ is the gas pressure and $p_r=aT^4/3$ is 
the radiation pressure. In the radiation pressure dominated
case $\beta\ll 1$, while in the gas pressure dominated case 
$1-\beta\ll 1$.

Behavior of viscosity in the radiation pressure dominated 
fluid is not well 
understood at the moment. There is still an ongoing debate 
whether it is determined by the full pressure or just the gas
pressure in the disk. For that reason we use a prescription 
motivated by the conventional $\alpha$-parametrization of 
Shakura \& Sunyaev (1973) and accounting for the two 
possibilities in a convenient form (Goodman 2003):
\ba
\nu=\alpha\beta^b \frac{c_s^2}{\Omega}.
\label{eq:nu}
\ea
Here $b=0$ corresponds to kinematic viscosity proportional 
to the full pressure $p$, while $b=1$ corresponds to the 
case when only the gas pressure $p_g$ determines the viscosity.
Lightman \& Eardley (1974) suggested that radiation
pressure dominated disks with $b=0$ are thermally unstable 
but recent numerical work by Hirose \etal (2009a,b) does not
show this to be the case.

Disks around SMBH binaries are heated by internal viscous 
dissipation and also by energy deposition of the density waves
launched by the central binary. For simplicity we do not consider 
here the latter contribution (its role has been studied by 
Lodato \etal 2009) and assume that disk heating is 
described by equations (\ref{eq:nrg}) and (\ref{eq:T_gen}).
Using equation (\ref{eq:Tscale}) we estimate the effective 
temperature of the disk as
\ba
T_{\rm eff}(r)&=&\left(\frac{3}{8\pi}
\frac{F_J\sqrt{GM_c}}{\sigma}\right)^{1/4}
r^{-7/8}
\label{eq:Teff_estimate_SMBH}
\\
&\approx &
1.1\times 10^{3}\mbox{K}~F_{J,50}^{1/4}M_{c,7}^{1/8}
r_{-2}^{-7/8},
\nonumber
\ea
where $M_{c,n}\equiv M_c/(10^nM_\odot)$, 
$r_{n}\equiv r/(10^{n}\mbox{pc})$, and $F_{J,n}\equiv
\FJ/(10^{n}\mbox{erg})$. 
This estimate assumes a particular value of $\FJ$, which we 
motivate in \S \ref{sect:F_J}.

In our treatment of the vertical radiation transfer in the disk 
we follow Goodman (2003) and relate the midplane disk 
temperature $T$ to $T_{\rm eff}$ via 
\ba
T^4=\frac{\tau}{2}T_{\rm eff}^4,
\label{eq:rad_transfer}
\ea
typical for optically thick ($\tau\gg 1$) disks, 
where $\tau=\kappa\Sigma$ is the optical depth. Also,
\ba
\Sigma=2\rho h,
\label{eq:surfdens}
\ea
where $\rho$ is the characteristic disk density, $h=c_s/\Omega$
is the disk scale height and $c_s=(p/\rho)^{1/2}$ is the sound 
speed, determined by the total pressure $p$. In the 
innermost region of the disk where the transition between the 
radiation and gas pressure dominated regimes occurs we 
assume opacity to be dominated by electron scattering, i.e. 
$\kappa=\kappa_{es}\approx 0.4$ cm$^2$ g$^{-1}$. Further out, 
in the gas pressure dominated part of the disk, $\kappa$ is
determined by the free-free opacity. We separately consider all
these regimes below. 

Combining equations (\ref{eq:Fnu}), (\ref{eq:T_gen}), (\ref{eq:nu}),
(\ref{eq:rad_transfer}), \& (\ref{eq:surfdens}) one finds the 
following equation determining the value of $\beta$ for a disk 
with opacity dominated by electron scattering:
\ba
\frac{\beta^{4+b}}{(1-\beta)^{10}}=2^6\pi^8
\left(\frac{k}{\mu}\right)^{4}\frac{c^{10}}
{\kappa_{es}^9\sigma}\frac{(GM_c)^{1/2}}{\alpha\FJ^8}r^{29/2}.
\label{eq:beta}
\ea
From that one can find the distance $r^{\rm rad/gas}$ of the transition 
between the radiation and gas pressure dominated regimes by 
setting $\beta=1/2$ in equation (\ref{eq:beta}):
\ba
r^{\rm rad/gas}&=&\left[\frac{2^b}{\pi^8}\alpha
\left(\frac{\mu}{k}\right)^{4}\frac{\kappa_{es}^9\sigma}{c^{10}}
\frac{\FJ^8}{(GM_c)^{1/2}}\right]^{2/29}
\label{eq:rad_tran}
\\
&\approx & 
3.8\times 10^{-3}\mbox{pc}~\left[2^{b}
\frac{\alpha_{-1}\mu_{0.5}^4 F_{J,50}^8}{M_{c,7}^{1/2}}
\right]^{2/29},
\nonumber
\ea
where $\alpha_n\equiv \alpha/10^n$, 
$\mu_n\equiv nm_p$, i.e. $\mu$ in this formula is normalized
by the molecular weight of fully ionized H. Note the 
extremely weak dependence of $r^{\rm rad/gas}$ on the binary mass
$M_c$.


\subsubsection{Radiation pressure dominated regime.}  
\label{sect:rad}

Interior to $r^{\rm rad/gas}$ disk is radiation pressure dominated
and by assumption $\kappa=\kappa_{es}$. In this case one 
finds
\ba
\frac{h(r)}{r}&=&\frac{1}{2\pi}\frac{\kappa_{es}}{c(1-\beta)}
\frac{\FJ}{\Omega r^3}
\label{eq:h_rad_es}
\\
&\approx &
1.1\times 10^{-3}~F_{J,50}M_{c,7}^{-1/2} r_{-2}^{-3/2},
\nonumber
\ea
irrespective of the value of $b$ in equation (\ref{eq:nu}). 
At the same time the scalings of $\Sigma(r)$ and $T(r)$ 
explicitly depend on the viscosity behavior: for $b=0$
\ba
\Sigma(r)&=&\frac{4\pi}{3}
\frac{c^2}{\alpha\kappa_{es}^2 F_J}r^2
\label{eq:sig_b=0}
\\
&\approx &
2.1\times 10^6\mbox{g cm}^{-2}~\alpha_{-1}^{-1}F_{J,50}^{-1} r_{-2}^2,
\nonumber
\\
T(r)&=&\left[\frac{(GM_c)^{1/2}c^2}{4\alpha\sigma\kappa_{es}}
\right]^{1/4}r^{-3/8}
\label{eq:T_b=0}
\\
&\approx &
2.9\times 10^4\mbox{K}~\alpha_{-1}^{-1/4}M_{c,7}^{1/8} r_{-2}^{-3/8},
\nonumber
\ea
while for $b=1$
\ba
\Sigma(r)&=&\left[\frac{2^4}{3^5\pi^3}
\left(\frac{\mu}{\alpha k}\right)^4
\frac{\sigma \FJ^3}{(GM_c)^{1/2}\kappa_{es}}\right]^{1/5}
r^{-9/10}
\label{eq:sig_b=1}
\\
&\approx &
6\times 10^4\mbox{g cm}^{-2}~\left[\frac{\mu_{0.5}^4 
F_{J,50}^3}{\alpha_{-1}^4 
M_{c,7}^{1/2}}\right]^{1/5}r_{-2}^{-9/10}
\nonumber
\\
T(r)&=&\left[\frac{1}{2^{4}\pi^2}
\frac{\mu}{k}\frac{(GM_c)^{1/2}\kappa_{es}}{\alpha\sigma}
\FJ^2\right]^{1/5}
r^{-11/10}
\label{eq:T_b=1}
\\
&\approx &
1.2\times 10^4\mbox{K}~\left[\frac{\mu_{0.5}M_{c,7}^{1/2}
F_{J,50}^2}{\alpha_{-1}}\right]^{1/5}r_{-2}^{-11/10}.
\nonumber
\ea

Note that a radiation pressure dominated disk with 
$\FJ=const$ (and $\dot M=0$) has $h(r)\propto r^{-1/2}$
(as opposed to the case of a conventional $\dot M=const$ 
disk for which $h(r)=const$), i.e. the disk {\it puffs up} as
$r$ decreases. Surface density shows dramatically different 
behavior depending on the value of $b$: it rises with 
$r$ when viscosity is proportional to the total pressure ($b=0$),
but drops with $r$ when $\nu$ scales with the gas pressure.
Finally, $T(r)$ is independent of $\FJ$ when $b=0$ but scales
as $\propto \FJ^{2/5}$ for $b=1$.


\subsubsection{Gas pressure dominated regime with 
$\kappa=\kappa_{es}$.}  
\label{sect:gas_es}

Outside of $r^{\rm rad/gas}$ gas pressure dominates, 
meaning that $\beta\to 1$ and $\nu\to \alpha c_s^2/\Omega$, 
but initially opacity is still determined by electron 
scattering, so that $\kappa=\kappa_{es}$. It is easy to see 
that in this case the midplane temperature and surface density 
runs in the disk should be the same as in the radiation pressure
dominated case with $b=1$: derivation of $T(r)$ 
and $\Sigma(r)$ involves only the equation of vertical radiation 
transfer (\ref{eq:rad_transfer}) and the definition (\ref{eq:Fnu}), 
which are identical in two cases because the viscosity and opacity 
are the same when $b=1$. Thus, the behavior of $\Sigma(r)$ and $T(r)$ 
in the gas pressure dominated case with $\kappa=\kappa_{es}$ is
given by equations (\ref{eq:sig_b=1}) and (\ref{eq:T_b=1}) 
correspondingly. 

However, the scaling of the aspect ratio of the disk with $r$
is different from $h/r$ given by equation (\ref{eq:h_rad_es}) 
since unlike the 
case studied in \S \ref{sect:rad} the vertical support is 
now provided by the gas pressure. As a result one finds
\ba
\frac{h(r)}{r}&=&\left[\frac{1}{16\pi^2}
\left(\frac{k}{\mu}\right)^4
\frac{\kappa_{es}\FJ^2}{\alpha\sigma (GM_c)^{9/2}}\right]^{1/10}
r^{-1/20}
\label{eq:hr_gas_es}
\\
&\approx &
6.6\times 10^{-3}\left[\frac{F_{J,50}^2}{\alpha_{-1}\mu_{0.5}^4
M_{c,7}^{9/2}}\right]^{1/10}
r_{-2}^{-1/20}.
\nonumber
\ea
Note that $h/r$ goes down with $r$ meaning that the disk is not 
flared. However, this dependence on $r$ is so weak that the 
aspect ratio is essentially constant with radius.


\subsubsection{Gas pressure dominated regime with 
$\kappa=\kappa_{ff}$.}  
\label{sect:gas_ff}

At even larger distances opacity in the disk is dominated 
by the free-free opacity $\kappa_{ff}=\kappa_{ff,0}\rho T^{-7/2}$,
with $\kappa_{ff,0}=8\times 10^{22}$ cm$^{-1}$ g$^{-2}$ K$^{7/2}$.
Using equations (\ref{eq:sig_b=1}) applicable for the situation 
described in \S \ref{sect:gas_es} and the behavior of the 
midplane temperature inferred from equation (\ref{eq:hr_gas_es})
one finds the transition between the two opacity regimes to take 
place at 
\ba
r^{\rm es/ff}&=&\left(\frac{3}{8\pi}\frac{\kappa_{es}^2 \FJ}
{\kappa_{ff,0}\sigma}\right)^{1/2}
\left(\frac{k}{\mu}\right)^{1/4}
\label{eq:r_es_ff}
\\
&\approx &
8.4\times 10^{-3}\mbox{pc}~F_{J,50}^{1/2}\mu_{0.5}^{-1/4}.
\nonumber
\ea
Outside this radius parameters of the disk scale as 
\ba
\Sigma(r)&=&\left[\frac{2^{10}}{3^{18}
\pi^{14}}\left(\frac{\sigma}{\kappa_{ff,0}}\right)^2
\left(\frac{\mu}{k}\right)^{15}
\frac{\FJ^{14}}{\alpha^{16}
(GM_c)^{2}}\right]^{1/20}
\label{eq:sig_ff}
\\
&\times & r^{-11/10}\approx 
7.2\times 10^4\mbox{g cm}^{-2}~\frac{\mu_{0.5}^{3/4} F_{J,50}^{7/10}}
{\alpha_{-1}^{4/5}M_{c,7}^{1/10}}
r_{-2}^{-11/10},
\nonumber
\\
T(r)&=&\left[\frac{2^{-5}}{3\pi^{3}}\frac{\kappa_{ff,0}}{\sigma}
\left(\frac{\mu}{k}\right)^{5/2}
\frac{\FJ^{3}(GM_c)}{\alpha^2}
\right]^{1/10}r^{-9/10},
\label{eq:T_ff}
\\
&\approx &
10^4\mbox{K}~\frac{\mu_{0.5}^{1/4}F_{J,50}^{3/10}M_{c,7}^{1/10}}
{\alpha_{-1}^{1/5}}r_{-2}^{-9/10},
\nonumber
\\
\frac{h(r)}{r}&=&\left[\frac{2^{-5}}{3\pi^{3}}
\frac{\kappa_{ff,0}}{\sigma}\left(\frac{k}{\mu}\right)^{15/2}
\frac{\FJ^{3}}{\alpha^{2}(GM_c)^{9}}
\right]^{1/20}r^{1/20},
\label{eq:hr_ff}
\\
&\approx &
6\times 10^{-3}\frac{F_{J,50}^{3/10}}
{\alpha_{-1}^{1/5}\mu_{0.5}^{3/4}M_{c,7}^{9/10}}
r_{-2}^{1/20},
\nonumber
\ea
see equations (\ref{eq:sig_general})-(\ref{eq:hr_general}) in 
Appendix \ref{sect:gen_opacity}. In this opacity regime disk 
is only weakly flared and $h/r$ is essentially constant 
with radius.


\subsubsection{Characteristic values of $\FJ$.}  
\label{sect:F_J}

We now try to motivate the characteristic values of the 
viscous angular momentum flux in disks around SMBH 
binaries using different arguments. In the case of 
constant $\dot M$ disks a standard way of providing the
characteristic value of $\dot M$ is through the Eddington 
mass accretion rate
\ba
\dot M_{\rm Edd}=\frac{4\pi GM_c}{c\kappa_{es}}\varepsilon^{-1}
\approx 0.2~M_\odot~\mbox{yr}^{-1}\frac{M_{c,7}}
{\varepsilon_{0.1}},
\label{eq:dotM_Edd}
\ea 
where $\varepsilon=0.1\varepsilon_{0.1}$ is the radiative 
efficiency.

In the case of constant $\FJ$ (or $\dot M=0$) disk one can
also formulate the Eddington limit on the value of $\FJ$, 
which is reached 
when $h/r\sim 1$. When this happens radiation pressure puffs 
up the disk to such extent that it becomes geometrically 
thick. Further increase of $\FJ$ should result in mass loss 
caused by the radiation pressure driven wind.
Another possibility for $h/r\sim 1$ is the mass overflow 
across the orbit of the secondary which may become possible 
since high above the midplane the potential of the secondary 
is weaker, potentially allowing the gas to cross the orbit 
of the secondary high above the disk midplane\footnote{For 
$q\ll 1$ puffing up of the disk may cause the overflow of
the secondary orbit even when $h/r\lesssim 1$.}. Both these 
processes result in mass loss for the circumbinary disk 
at its inner edge, which can have important implications 
for the orbital evolution of the central binary as we 
demonstrate in \S \ref{sect:evol}.

Unlike the situation with $\dot M_{\rm Edd}$ in conventional 
constant $\dot M$ disks, the Eddington limit on $\FJ$ found 
from the condition $h/r=1$ depends on the value of $r$ at 
which it is evaluated, see equation (\ref{eq:h_rad_es}):
\ba
\FJE(r)=2\pi\frac{(GM_c)^{1/2}c}{\kappa_{es}}
r^{3/2}.
\label{eq:FJ_Edd}
\ea
It is clear that this condition is most constraining at the 
inner edge of the disk $r_{in}\sim r_b$, where $r_b$ is the 
semi-major axis of the SMBH binary. 

Obviously, one needs to invoke additional considerations to 
pick a particular value of $r_{in}$ (or $r_b$) at which 
$\FJE$ is to be evaluated. Here we assume that 
this critical value of $r_{in}$ is such that at $r_b\approx r_{in}$ 
the orbital decay timescale of the binary due to emission 
of gravitational waves $t_{\rm GW}$ is equal to the 
characteristic timescale on which the orbit of the 
binary shrinks due to the tidal coupling to the circumbinary 
disk $t_J$. The logic behind choosing this condition 
is that the disk then stays sub-Eddington all the way until 
the point when the 
GW emission becomes more important for the 
orbital evolution of the binary than its tidal coupling 
to the disk. Beyond this point the disk becomes super-Eddington 
and starts losing mass in a radiation pressure driven wind, 
reducing the disk torque. However this does not affect the 
binary inspiral since the orbital evolution 
is no longer sensitive to the tidal torque.

Equating
\ba
t_{\rm GW}(r_b)=\frac{5}{2q_S}\frac{R_S}{c}
\left(\frac{r_b}{R_S}\right)^4
\label{eq:t_GW}
\ea
(here $q_S\equiv 4q/(1+q)^2$ and
$R_S\equiv 2GM_c/c^2$ is the Schwarzschild radius of the black 
hole with the mass equal to the total mass of the binary 
$M_c$) and
\ba
t_J(r_b)=\left|\frac{d\ln r_b}{dt}\right|^{-1}=\frac{L_b}{2\FJ}=
\frac{q_S}{8}\frac{M_c\sqrt{GM_c r_b}}{\FJ}
\label{eq:t_J}
\ea 
(here $L_b=(q_S/4)M_c(GM_c r_b)^{1/2}$ is the total 
orbital angular momentum of the binary) one finds 
\ba
r_{in}=R_S\left(\frac{q_S^2}
{2^{5/2}5}\frac{M_c c^2}{\FJ}\right)^{2/7}.
\label{eq:r_in_crit}
\ea
Plugging this value of $r_{in}$ into equation (\ref{eq:FJ_Edd})
one finds that the Eddington value of the viscous angular momentum 
flux in $\dot M=0$ disk based on the condition 
$t_{\rm GW}(r_{in})=t_J(r_{in})$ is given by
\ba
\FJE &=&\left[\frac{2^{10}3^3 \pi^7}{5^3}
\frac{(GM_c)^{14}M_c^3q_S^6}{c^8\kappa_{es}^7}\right]^{1/10}
\label{eq:FJEdd}
\\
&\approx &
10^{51}\mbox{erg}~M_{c,7}^{17/10}q_S^{3/5}.
\ea
This argument justifies the adoption of a characteristic value
of $\FJ=10^{50}$ erg (corresponds to $M_c=10^7$ M$_\odot$, 
$q=0.005$) in our numerical estimates.

One can come up with other ways of choosing critical $r_{in}$
or the characteristic value of $\FJ$. In particular, one may 
demand the disk to stay sub-Eddington until the point when 
$t_{\rm GW}$ becomes equal to the viscous timescale at $r_{in}$,
after which the binary orbit shrinks faster than the viscosity 
can refill the central cavity (Milosavljevi\'c \& Phinney 2005). 
This happens at considerably smaller value of $r_{in}$ than 
that given by (\ref{eq:r_in_crit}) and implies lower 
$\FJ$. However, (1) at this point tidal torque is already 
completely negligible compared to the angular momentum loss 
due to the GW radiation, and (2) such condition 
would depend on the poorly understood value of $b$ in equation
(\ref{eq:nu}). Thus, we avoid this way of constructing Eddington 
limit-based estimate of $\FJ$.

One can also evaluate $\FJ$ based on arguments completely 
independent of the Eddington limit. For example, one may 
demand the value of $t_{\rm GW}$ at $r_{in}$ given by 
(\ref{eq:r_in_crit}) to be equal to some characteristic 
time $t$. This would imply that after the GW 
emission becomes the dominant cause of the binary inspiral,
the lifetime of the binary until merger is equal to $t$. 
Such estimate gives
\ba
F_{J,t} &=& M_c c^2\frac{q_S^{9/8}}{2^{27/8}5^{1/8}}
\left(\frac{R_S}{ct}\right)^{7/8}
\label{eq:FJ_t}
\\
&\approx &
4\times 10^{46}\mbox{erg}~q_S^{9/8}M_{c,7}^{15/8}
\left(\frac{t}{10^{10}\mbox{yr}}\right)^{-7/8}.
\nonumber
\ea
It sets a lower limit on the value of $\FJ$
necessary for equal mass ($q_S=1$) $M_c=10^7$ M$_\odot$ 
SMBH binary to merge within the Hubble time. For 
$\FJ=\FJE$ such a binary would merge within 
$4\times 10^5$ yr after the GW emission starts 
to dominate its orbital evolution.

\begin{figure}
\plotone{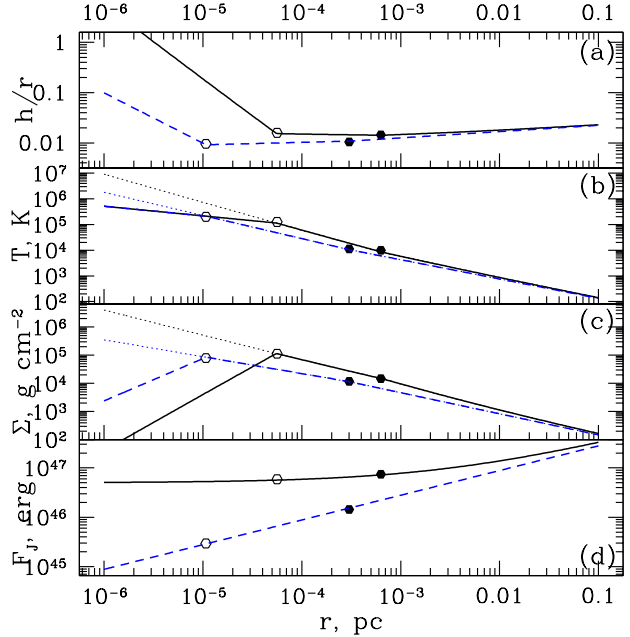}
\caption{
Properties of a steady disk around a SMBH binary 
with mass $M_c=10^5$ M$_\odot$ described by a solution
(\ref{eq:Fsol}) with $F_{J,0}=5\times 10^{46}$ erg and 
mass accretion rate at infinity 
$F_{J,1}=\dot M_\infty=\dot M_{\rm Edd}$ (radiative efficiency 
of $0.1$ is assumed). Solid (black) curves describe the run of (a) aspect 
ratio $h/r$, (b) temperature $T$, (c) surface density $\Sigma$, 
and (d) angular momentum flux $F_J$ with $r$ for a disk with 
$b=0$ in radiation pressure dominated part of the disk. Dashed 
(blue) curves describe the same but for a standard constant $\dot M$ 
disk with the same mass accretion rate $\dot M_\infty$. Dotted 
extensions of these curves at small $r$ correspond to $b=1$ in 
radiation pressure dominated part of the disk. Open dots on each
curve correspond to the transition between the radiation and
gas pressure dominated regimes, while filled dots describe the
transition between the electron scattering 
and  free-free opacity.
\label{fig:properties}}
\end{figure}

Finally, using equations (\ref{eq:Fnu}) and (\ref{eq:nu}) one 
can also express $\FJ$ via the disk mass enclosed
between its inner edge and some outer radius $r_o$ via
\ba
M_{disk}=\frac{2\FJ}{3\alpha}\int\limits_{r_{in}}^{r_o}
\frac{dr}{rc_s^2}\sim \frac{\FJ}{\alpha c_s^2(r_o)},
\label{eq:M_d}
\ea
where it is assumed that gas pressure dominates 
(or $b=1$), $c_s=(kT/\mu)^{1/2}$) and 
$\FJ=const$ between $r_{in}$ and $r_o$. 
The approximate relation in this formula is valid for 
$r_o\gg r_{in}$, provided that the outer regions of the disk 
dominate its mass --- a rather 
natural assumption as long as the midplane temperature
falls with increasing radius, as this equation shows. 
Note that $M_{disk}$ explicitly depends only on the 
disk temperature at the outer radius but not $r_o$ itself.

Despite the estimates (\ref{eq:FJ_t}) and (\ref{eq:M_d}) which 
have clear physical 
meaning, we still advocate the use of $\FJE$ as it represents an
important upper limit on $\FJ$: for $\FJ\lesssim \FJE$ the
orbital evolution of the binary is essentially not affected
by the Eddington limit (even though at late stages of inspiral
the inner part of the disk may become super-Eddington and be
depleted by the radiation pressure driven wind, see \S 
\ref{sect:Edd_lim}).


\subsubsection{Global properties of steady state disks.}  
\label{sect:global_properties}

Results of \S\S \ref{sect:rad}-\ref{sect:gas_ff} allow us to 
understand global characteristics of steady circumbinary disks 
described by the solution (\ref{eq:Fsol}).  

In Figure \ref{fig:properties} we demonstrate how the main
properties of such a disk vary as a function of $r$ across 
regions with different opacity and pressure behavior. We 
consider a $M_c=10^5$ M$_\odot$ SMBH binary orbited by a disk 
described by the relation  (\ref{eq:Fsol}) with 
$F_{J,0}=5\times 10^{46}$ erg and 
$F_{J,1}=\dot M_\infty=\dot M_{\rm Edd}$. Far from the binary 
(at $r\to \infty$) disk transitions to a standard constant 
$\dot M=\dot M_\infty$ accretion disk, while at small 
separations corresponding to 
$l\lesssim F_{J,0}/\dot M_\infty$ it becomes a constant
$\FJ$ disk. Note that $\dot M=\dot M_\infty$
at all radii meaning that mass has to be removed at $r_{in}$ 
at the same rate with which it is supplied at large 
radii.

We compare the properties of this disk to a standard 
constant $\dot M=\dot M_\infty$ disk in which $F_{J,0}=0$
(dashed curves). We find that the midplane temperature $T$ is 
higher in a disk with non-zero $F_{J,0}$ in all regimes. 
As a result, this disk is more extended 
vertically and becomes geometrically thick ($h/r\sim 1$) 
at $r=3\times 10^{-6}$ pc ($\approx 300$ R$_S$) in 
the radiation pressure dominated regime. Unless the disk is 
truncated on the inside by the binary torque at larger separation, 
it would 
presumably be losing mass at this point. Another consequence 
of higher $T$ is that all transitions between different 
regimes occur at larger $r$ (typically by a factor of 
several) in the disk with non-zero $F_{J,0}$. In particular,
radiation pressure starts to dominate in this disk at
$\approx 5\times 10^{-5}$ pc as opposed to 
$\approx 10^{-5}$ pc in the $F_{J,0}=0$ disk. 

The comparison of the surface density structure between the 
two disks depends on whether $b=1$ or $0$ in the radiation 
pressure dominated regime. In the former case $\Sigma$
keeps increasing towards small $r$ in both types of disks, 
which reflects the inefficiency of viscosity proportional to
(relatively small) gas pressure in the $b=1$ case.
We find $\Sigma$ to be higher in the $F_{J,0}\neq 0$ disk 
by an order of magnitude in the radiation pressure 
dominated regime.  

In the case of $\nu$ scaling with the radiation pressure 
($b=0$) the behavior of $\Sigma$ is completely different
as it falls towards small $r$, because of rapid increase of 
$T$ and $\nu$ in the radiation pressure dominated regime.
As a result, close to the binary surface density is much 
smaller (by more than an order of magnitude) in the 
disk with non-zero $F_{J,0}$. The relatively small amount 
of mass residing in the vicinity of the SMBH binary in 
this case may have implications for the 
properties of the afterglow following the binary merger.

It is well known (Goodman 2003) that the outer parts of 
disks around SMBHs may be subject to 
gravitational instability (Safronov 1960; Toomre 1964; 
Goldreich \& Lynden-Bell 1965) 
when the Toomre Q parameter 
defined as $Q=\Omega c_s/(\pi G\Sigma)$ drops below unity. 
In the case of constant $\FJ$ disks or steady disks with 
$\FJ(l)$ given by the solution (\ref{eq:Fsol}) one can easily
determine the radius $r^{sg}$ at which the disk becomes 
self-gravitating using our results for $T(r)$ and $\Sigma(r)$ 
as a function of $\FJ$ 
derived in \S \ref{sect:rad}-\ref{sect:gas_ff} for 
different physical regimes. In the interest of 
brevity we do not perform this straightforward exercise here.
For constant $\dot M$ disks described by the solution 
(\ref{eq:Fsol}) with $F_{J,0}=0$ 
corresponding expressions for $r^{sg}$ can be found in 
Goodman (2003) and Haiman \etal (2009). Other effects that
may invalidate our treatment at large separations --- low
optical depth of the disk when $\Sigma$ becomes small,
neutrality of the disk at low $T$, etc. --- have been 
previously discussed in Haiman \etal (2009).


\section{Disk evolution.}  
\label{sect:evol}

Steady state solutions of the master equation (\ref{eq:evF})
discussed so far require rather special circumstances to
be realized globally, such as $\dot M$ which is independent of $r$.
In real circumbinary disks this condition is difficult
to realize, simply because the mass supply rate of the disk
at large separations $\dot M_\infty$ is determined by 
processes that have nothing to do with the central binary. 
At the same time $\dot M$ in the inner disk is set by the 
binary torque, and is in general different from 
$\dot M_\infty$. 

Because of this mismatch of $\dot M$ in different parts of the 
disk gas has to accumulate somewhere  and this naturally leads 
to the evolution of disk properties. As we will see later evolution
typically leads to the establishment of the quasi-steady state 
in the inner parts of the disk, where one can then apply the 
results obtained in previous section. To understand this 
process we need to obtain the time-dependent solutions of equation 
(\ref{eq:evF}). This necessarily requires specifying the 
dependence of the diffusion coefficient $D_J$ given by equation 
(\ref{eq:D_J}) upon $l$ and $\FJ$ and a set of boundary 
conditions. 

With a rare exceptions discussed in \S \ref{sect:general} we 
will always assume that the circumbinary disk starts out as a 
conventional constant $\dot M$ disk in which 
\ba
\FJ(t=0,l)=\dot M_\infty l,
\label{eq:init_sol}
\ea
see equations (\ref{eq:Mdot_SS}) and (\ref{eq:dotM0}). This is 
a natural assumption for a viscously evolving disk, which is 
supplied by a source of mass located at $r\gg r_b$. Indeed, 
as the mass flows in, the viscous time in the disk 
decreases with $r$ causing its structure to converge 
to that of a standard constant $\dot M$ disk. 
When the inner edge of the 
disk approaches the semi-major axis of the binary, 
the latter starts tidally interacting with the disk providing 
a source of angular momentum. Tidal 
torque on the disk rapidly increases and stops the mass
inflow at some radius $r_{in}$ (or, equivalently, some value of
the specific angular momentum $l_{in}$), which is comparable to
the semi-major axis of the binary. We consider evolution of the 
disk starting from this moment and use an inner boundary 
condition in the form 
\ba
\frac{\partial\FJ}{\partial l}\Big|_{l=l_{in}}=\dot M(l_{in})=
\chi\dot M_\infty,
\label{eq:inner_BC}
\ea
see equation (\ref{eq:inner_BC0}).

\begin{figure}
\plotone{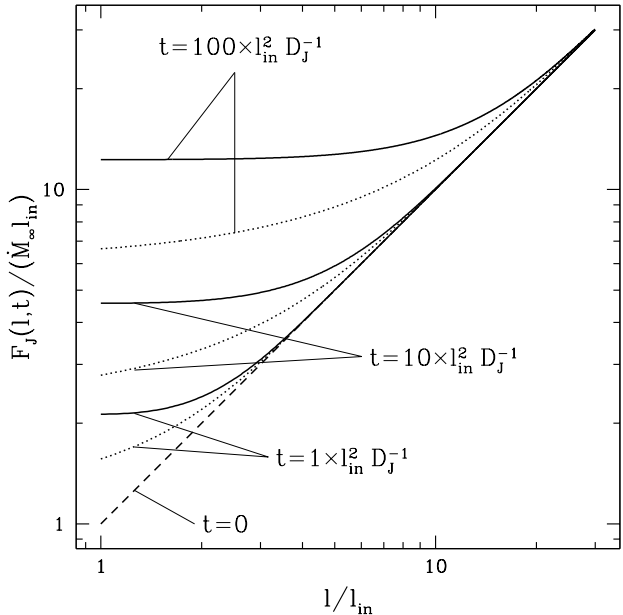}
\caption{
Time evolution of the angular momentum flux distribution $F_J(l,t)$
($l$ is the specific angular momentum acting as a space-like 
coordinate) in a disk with constant $D_J$ (see Eq. 
[\ref{eq:const_D_J_sol}]) and different boundary conditions 
imposed at the inner disk radius $r_{in}$, 
at which the specific angular momentum is $l_{in}$: $\dot M(l_{in})=0$
({\it solid lines}) and $\dot M(l_{in})=0.5\dot M_\infty$
({\it dotted lines}). The initial distribution of 
$F_J(l,0)=\dot M_\infty l$ assumes a standard constant 
$\dot M=\dot M_\infty$ disk and is shown by a dashed line. 
Snapshots of $F_J$ at different moments of time (labeled on
the Figure) are shown. 
\label{fig:linear}}
\end{figure}

In some cases one can solve equation (\ref{eq:evF}) 
analytically, which is useful for qualitative understanding
of the more complicated situations. In particular, for
$D_J=const$ with initial and boundary conditions specified 
by equations (\ref{eq:init_sol}) and (\ref{eq:inner_BC}) 
correspondingly one finds the following solution (with $l$ playing 
role of spatial coordinate): 
\ba
\FJ(t,l)&=&\chi\dot M_\infty l+(1-\chi)\dot M_\infty
\nonumber
\\
&\times &\left\{
l_{in}+(l-l_{in})\mbox{erf}\left(\frac{l-l_{in}}{2\sqrt{D_J t}}\right)
\right.
\nonumber
\\
&+&\left. \left(\frac{4D_Jt}{\pi}\right)^{1/2}\exp
\left[-\frac{(l-l_{in})^2}{4D_J t}\right]\right\}.
\label{eq:const_D_J_sol}
\ea

This solution is shown in Figure \ref{fig:linear} for different 
values of $\dot M(l_{in})$ at different moments of time. One
can clearly see that as the time goes by the influence
of the central binary extends to larger and larger values of 
$l$, implying also larger distance from the central binary $r$. 
The transition from the initial distribution of $F_J$ 
given by equation (\ref{eq:init_sol}) to $\FJ$ strongly 
affected by the binary torque occurs at\footnote{Here we 
follow the notation of IPP, who called the radius
at which this transition occurs the ``radius of influence''
$r_{\rm infl}$.}  $l=l_{\rm infl}
\sim (D_Jt)^{1/2}$.

Note that the solution for $\dot M(l_{in})=0$ ($\chi=0$) 
clearly exhibits an inner region with $\FJ\to const$, in 
agreement with our discussion of $\dot M=0$ solutions in 
\S \ref{sect:steady}. At the same time the solution for 
$\dot M(l_{in})=0.5\dot M_\infty$ ($\chi=0.5$) develops 
an inner region with $\FJ$ linearly increasing with $l$,
again in complete agreement with the solution (\ref{eq:Fsol})
with $F_{J,1}\neq 0$. Note that the torque exerted on the
disk by the binary $\FJ(l_{in})$ is quite different in two 
cases: it is smaller for larger values of $\chi$ which is 
natural since higher $\chi$ implies less mass accumulation 
at the inner edge of the disk and less torque exerted by 
the binary on the disk. This means that higher $\chi$
(higher $\dot M(l_{in})$) should result in slower inspiral
of the binary. 

We show next that the main features of disk evolution 
illustrated for the $D_J=const$ remain valid for
the more general behavior of $D_J$. Following Filipov 
(1984), Lyubarskij \& Shakura (1987), Pringle (1991), IPP,
and Lipunova \& Shakura (2000) we now derive self-similar 
solutions for evolving structure of an externally fed 
and centrally torqued circumbinary disk, allowing for 
the possibility of some mass overflow across the orbit 
of the secondary. Based on these solutions we 
then outline in \S \ref{sect:general} a general picture 
of the non-self-similar disk evolution.


\subsection{Self-similar solutions for circumbinary disk evolution.}  
\label{sect:self}

In \S \ref{sect:properties} we demonstrated that in a 
variety of situations disk properties can be expressed
as power laws of different physical parameters ---
$\FJ$, $r$ (or, equivalently, $l$), $M_c$, etc. Based on 
that we show in Appendices \ref{sect:gen_opacity} \&
\ref{sect:D_J} that the diffusion coefficient in different 
regimes can be generically expressed in terms of the angular 
momentum flux $\FJ$ and specific angular momentum $l$ in the 
power law form 
\ba
D_J=D_{J,0}~\FJ^d ~l^p,
\label{eq:PL_diff}
\ea
where $d$ and $p$ are constant power law indices. The explicit 
expressions for $D_{J,0}$, $d$, and $p$ in relevant regimes
can be found in Appendix \ref{sect:D_J}. 

Equation (\ref{eq:evF}) with $D_J$ in the form (\ref{eq:PL_diff})
admits a self-similar solution\footnote{Previously Lyubarskij \& 
Shakura (1987) derived such solutions for $\FJ$ assuming boundary 
conditions different from what we use here.} provided that the 
problem at hand
has no intrinsic scale. In circumbinary disks the inner edge of
the disk (comparable to the semi-major axis of the binary) sets
a natural scale. However, after the angular momentum injected 
by the binary has been transmitted by viscosity out to distances 
large compared to the radius of the inner cavity, this scale 
should not affect system's behavior and
evolution becomes self-similar. 

To illustrate this 
point let us consider the solution (\ref{eq:const_D_J_sol}) 
corresponding to a particular case of $d=p=0$ in the limit of 
$t\gg l_{in}^2/D_J$, so that $l_{\rm infl}\gg l_{in}$ 
(or $r_{\rm infl}\gg r_{in}$). In this limit we can rewrite 
the solution (\ref{eq:const_D_J_sol}) as
\ba
\FJ(t,l)&=&\dot M_\infty\sqrt{D_J t}\left[\chi\xi+(1-\chi)
\xi\mbox{erf}\left(\frac{\xi}{2}\right)\right.
\nonumber
\\
&+&\left.(1-\chi)\frac{2}{\sqrt{\pi}}\exp
\left(-\frac{\xi^2}{4}\right)\right].
\label{eq:const_D_J_sol_self_sim}
\ea
where $\xi=l/\sqrt{D_Jt}$ is the dimensionless coordinate, which, as 
we will see later, plays the role of an independent self-similar 
variable. It is clear from this result that at late times 
the solution of the evolutionary equation (\ref{eq:evF}) 
for $D_J=const$ is 
indeed independent of the exact value of $l_{in}$ at which the 
inner boundary condition is imposed, and one can effectively 
set $l_{in}$ to zero thus eliminating any intrinsic scales from 
the problem at hand. 

In general case of arbitrary $d$ and $p$ in the expression 
(\ref{eq:PL_diff}) we first define new variables
\ba
f_J\equiv \frac{F_J}{\dot M_\infty},~~~
\tau\equiv D_{J,0}\dot M_\infty^d t,
\label{eq:var_change}
\ea 
transforming equation (\ref{eq:evF}) into
\ba
\frac{\partial}{\partial \tau} f_J^{1-d}=l^p
\frac{\partial^2 f_J}{\partial l^2},
\label{eq:evF_dim-less}
\ea 
with the boundary condition 
\ba
\frac{\partial f_J}{\partial l}\Big|_{l=0}=\frac{\dot M(l=0)}
{\dot M_\infty}=\chi,
\label{eq:inner_BC_dim-less}
\ea
(instead of equation [\ref{eq:inner_BC}]), where $\chi\le 1$
and may be equal to zero. 

\begin{figure}
\plotone{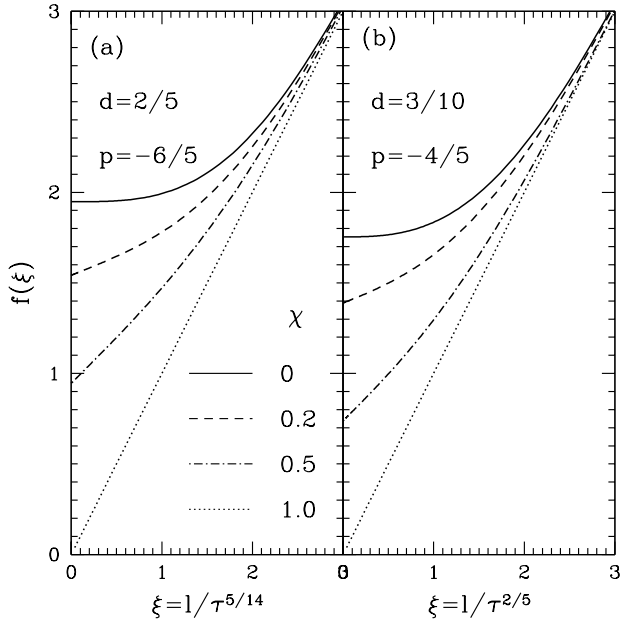}
\caption{
Behavior of the self-similar function $f$ vs the 
variable $\xi$ defined in equation (\ref{eq:self_sim_form}). Different
panels correspond to different values of the power law 
parameters $d$ and $p$: (a)
gas pressure dominated case with $\kappa=\kappa_{es}$, (b) 
gas pressure dominated case with $\kappa=\kappa_{ff}$. 
Various curves represent 
the run of $f(\xi)$ for different boundary conditions imposed
at the inner edge, allowing for the possibility of mass inflow 
past the secondary orbit: $\dot M(0)=\chi\dot M_\infty$.
\label{fig:self-similar}}
\end{figure}

One can easily see that equation (\ref{eq:evF_dim-less}) admits 
self-similar solutions in the form 
\ba
f_J=\tau^n f(\xi),~~~
\xi\equiv\frac{l}{\tau^n},~~~n=-\frac{1}{d+p-2},
\label{eq:self_sim_form}
\ea
where function $f$ satisfies ordinary differential
equation
\ba
f^{\prime\prime}f^d\xi^p=
n(1-d)(f-\xi f^\prime)
\label{eq:ODE}
\ea
with the boundary conditions
\ba
f^\prime(\xi\to 0)=\chi,~~~ f^\prime(\xi\to \infty)=1,
\label{eq:self-sim_BCs}
\ea
where the second boundary condition follows directly from 
the initial condition (\ref{eq:init_sol}). Note that for $d=p=0$ 
equation (\ref{eq:self_sim_form}) reproduces the similarity 
seen in the analytical solution (\ref{eq:const_D_J_sol_self_sim})
obtained for constant $D_J$. We also verified the 
self-similar scalings of $\Sigma$ derived in IPP for $\chi=0$
case.

In Figure (\ref{fig:self-similar}) we show the solutions of 
this equation for different values of mass accretion rate 
across the orbit of the secondary parametrized by the 
value of $\chi$ (generalizing results of IPP to the case of 
non-zero $\chi$). 
We have chosen two sets of $d$ and $p$ 
corresponding to astrophysically relevant situations:
gas pressure dominated regime with $\kappa=\kappa_{es}$ 
($d=2/5, p=-6/5$, $n=5/14$), and with $\kappa=\kappa_{ff}$
($d=3/10, p=-4/5$, $n=2/5$), see Appendix \ref{sect:D_J} or 
Lyubarskij \& Shakura (1987).
We will see later in \S \ref{sect:SMBHs} that the radius 
of influence often extends into the parts of the disk where 
one of these regimes is valid. 

As expected, the transition from the outer solution 
$f(\xi)=\xi$ unaffected by the binary torque to the inner 
solution influenced by it always occurs at 
$\xi\approx 1-2$. This transition clearly corresponds to the 
radius at which the viscous time in the disk $t_{visc}$ is equal to the 
evolution time $t$ of the system, i.e. the time that has passed 
since the central binary started tidally interacting with the 
disk at its inner edge. In other words, $t_{visc}(\xi\sim 1)\sim t$.

We take the radius of influence 
$r_{\rm infl}$ to correspond to 
$\xi_{\rm infl}=1$. Then, according to
equations (\ref{eq:var_change}) and (\ref{eq:self_sim_form}) 
the value of the specific angular momentum $l_{\rm infl}$ 
at $r_{\rm infl}$ is given by 
\ba
l_{\rm infl}(t)= \left(D_{J,0}\dot M_\infty^d\right)^n 
t^n,~~~r_{\rm infl}(t)=\frac{l_{\rm infl}^2(t)}{GM_c},
\label{eq:l_infl}
\ea 
while the viscous angular momentum flux in the disk is
\ba
F_J(l,t)=\dot M_\infty l_{\rm infl}(t)\times
f\left(l/l_{\rm infl}(t)\right),
\label{eq:FJ_self}
\ea
where the value of $f(\xi)$ can be found from Figure 
\ref{fig:self-similar} for a given $\chi$. 

\begin{figure}
\plotone{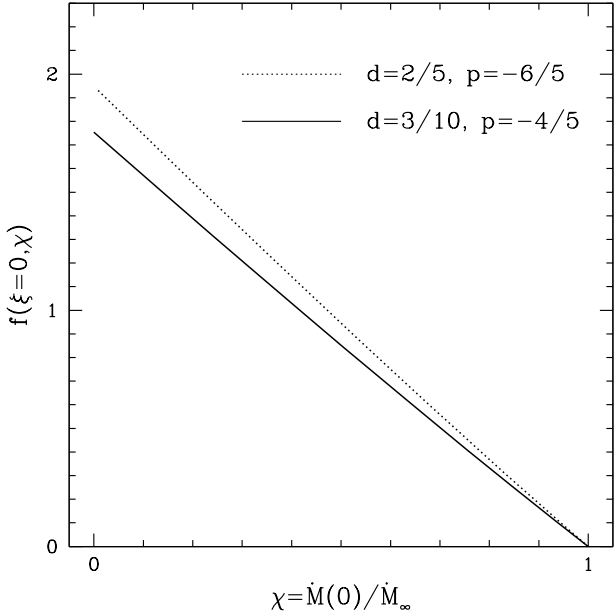}
\caption{
Value of $f(\xi=0,\chi)$ as a function of $\chi$ --- the 
fraction of $\dot M_\infty$ passing through the orbit of the 
secondary. 
Different curves correspond to different values of the power law 
parameters $d$ and $p$ indicated on panels: ({\it dotted}) 
gas pressure dominated case with $\kappa=\kappa_{es}$, 
({\it solid}) gas pressure 
dominated case with $\kappa=\kappa_{ff}$. Knowledge of $f(\xi=0)$
allows computation of the torque acting on the central binary for
arbitrary value of the ``accretion fraction'' $\chi$.
\label{fig:f0_chi}}
\end{figure}

The torque acting on the central binary in the limit 
$l_{in}\ll l_{\rm infl}(t)$ (or $r_{in}\ll r_{\rm infl}(t)$) 
is given simply by $F_J(0,t)=\dot M_\infty l_{\rm infl}(t)
f(0,\chi)$ where the dependence of $f(0,\chi)$ upon the 
efficiency $\chi$ of
mass inflow through the secondary orbit is shown in Figure  
\ref{fig:f0_chi} for different types of disks. One can see 
in complete analogy with the analytical solution 
(\ref{eq:const_D_J_sol}) that allowing for some mass flow 
across the orbit of the secondary (i.e. assuming non-zero 
$\chi$) leads to the reduction 
of the torque acting on the binary.

It might seem surprising that the torque experienced by 
the binary is independent of the mass of the secondary --- 
$F_J(0,t)$ is set only by $\dot M_\infty$ and $l_{\rm infl}(t)$, 
since it is the potential of the secondary that gives rise 
to the tidal coupling with the disk. We comment on this point 
in \S \ref{sect:SMBHs}.


\subsection{General description of the disk evolution.}  
\label{sect:general}

In \S \ref{sect:self} we outlined main features of the 
self-similar evolution of a circumbinary disk which arises
when the three essential conditions are met: (1) the behavior
of the diffusion coefficient $D_J$ is given by a simple power
law form (\ref{eq:PL_diff}), (2) the outer parts of the 
disk are well approximated 
by a standard constant $\dot M$ disk with $\FJ=\dot M_\infty l$,
and (3) the radius of influence $r_{\rm infl}$ far exceeds
the semi-major axis of the central binary $r_b$ (and the radius 
of the inner disk edge $r_{in}$). We now describe how the 
picture of the disk evolution changes when these assumptions
are relaxed by concentrating on the situation when there
is no mass inflow across the orbit of the secondary, i.e. 
$\chi=0$ or $\dot M(r_{in})=0$. More complicated setup allowing
for some mass inflow across the orbit of the secondary can
be understood by generalization of the picture that 
emerges in $\dot M(r_{in})=0$ case. 

Different parts of circumbinary disks can feature 
different physical regimes as illustrated in \S 
\ref{sect:properties}. In this case a simple form of
$D_J$ given by equation (\ref{eq:PL_diff}) will 
not work in the whole disk invalidating global 
self-similarity. However, $D_J$ can still be cast in 
this form in certain distance intervals with
power law indices $d$ and $p$ intrinsic to each region.
For example, one can easily imagine that the inner disk
is in the radiation pressure dominated regime, while 
further out it transitions to gas pressure dominated 
regime with the opacity initially given by $\kappa_{es}$ 
and then by $\kappa_{ff}$. In all these regimes
one can locally use the power law description of $D_J$ as 
described in Appendix \ref{sect:D_J}, with smooth 
transitions between the different scalings at the boundaries 
of different regimes. 

Evolution equation written in the form (\ref{eq:evF}) easily 
allows us to understand the behavior of the disk properties in this 
more complicated situation. Indeed, interior to $r_{\rm infl}$
the viscous time in the disk gets shorter with decreasing $r$,
meaning that for $r\lesssim r_{\rm infl}$ disk tends to approach
a steady state solution. Then equation (\ref{eq:evF}) implies that 
$\FJ$ in this part of the disk is given by a simple solution
(\ref{eq:Fsol}) {\it independent} of the complicated behavior
of $D_J$ caused by the transitions between different physical 
regimes. Disk properties such as $\Sigma(r)$, $T(r)$, etc. 
can be computed as functions of this
radially constant $\FJ$ (for $\dot M(r_{in})=0$) using 
formulae derived in \S 
\ref{sect:properties}, and will show different dependence 
on $r$, $\FJ$ and other system parameters in different regimes.

At every moment of time the value of the radially constant 
(for $r\lesssim r_{\rm infl}$) angular momentum flux $\FJ$ 
is obviously set by the the disk properties in a particular 
physical regime corresponding to $r\sim r_{\rm infl}$. 
This regime can change in time since $\FJ$ steadily 
increases and both the transition radii of different 
regimes and $r_{\rm infl}$ vary. Nevertheless, it is clear 
that our results for the self-similar disk evolution obtained
previously should allow one to easily understand even more 
complicated situations, see \S \ref{sect:SMBHs}.  

Second complication arises if the circumbinary disk does not 
start out as a standard constant $\dot M$ disk with the 
initial distribution of the angular momentum flux in the 
form (\ref{eq:init_sol}) but is characterized by some more 
complicated initial distribution of $\FJ(l,t=0)$. Again, our 
understanding of the self-similar disks allows us to 
qualitatively characterize disk evolution in this case. 
The region influenced by the binary torque would still 
expand in time with the dependence 
$l_{\rm infl}(t)$ given by an implicit relation
\ba
l_{\rm infl}^2\sim t\times D_J\left(\FJ(l_{\rm infl},t=0),
l_{\rm infl}\right),
\label{eq:gen_rel}
\ea
see equation (\ref{eq:evF}), in which we explicitly indicated 
the dependence of $D_J(\FJ,l)$ on $\FJ$ and $l$. Interior 
to $r_{\rm infl}$ angular momentum flux is roughly constant 
with radius and equal to $\FJ(l_{\rm infl}(t),t=0)$.

At the same time outside $r_{\rm infl}(t)$ the disk will 
still maintain the distribution of $\FJ$ close to the
initial distribution $\FJ(l,t=0)$ since the viscous time 
there is long compared to the system lifetime (which 
is also equal to $t_{visc}(r_{\rm infl})$). If the initial 
distribution of $\FJ$ exhibits a maximum at some radius 
$r_{max}$ then past the moment when $r_{\rm infl}\sim r_{max}$   
the circumbinary disk will turn into a decretion disk 
(Pringle 1991) and the mass accumulated in the central 
part of the disk will start flowing out, driven by  
continuing injection of the angular momentum by the binary  
(assuming that the binary does not merge by 
that time).

Finally, initially the radius of influence may not strongly
exceed the radius of the inner edge of the disk. This is true
if the evolutionary lifetime of the system has not yet
exceeded the viscous time at the inner edge of the disk. 
However, at later times the condition $r_{\rm infl}\gtrsim 
r_{in}$ is guaranteed to be fulfilled since $r_{\rm infl}$ 
steadily grows while both the binary semi-major axis 
and $r_{in}$ can only decrease. As a result, at late times 
the system should inevitably converge to the self-similar 
mode of evolution (see e.g. evolution shown in Figure 
\ref{fig:linear}) or its generalizations described above 
for the more complicated situations.

\begin{figure}
\plotone{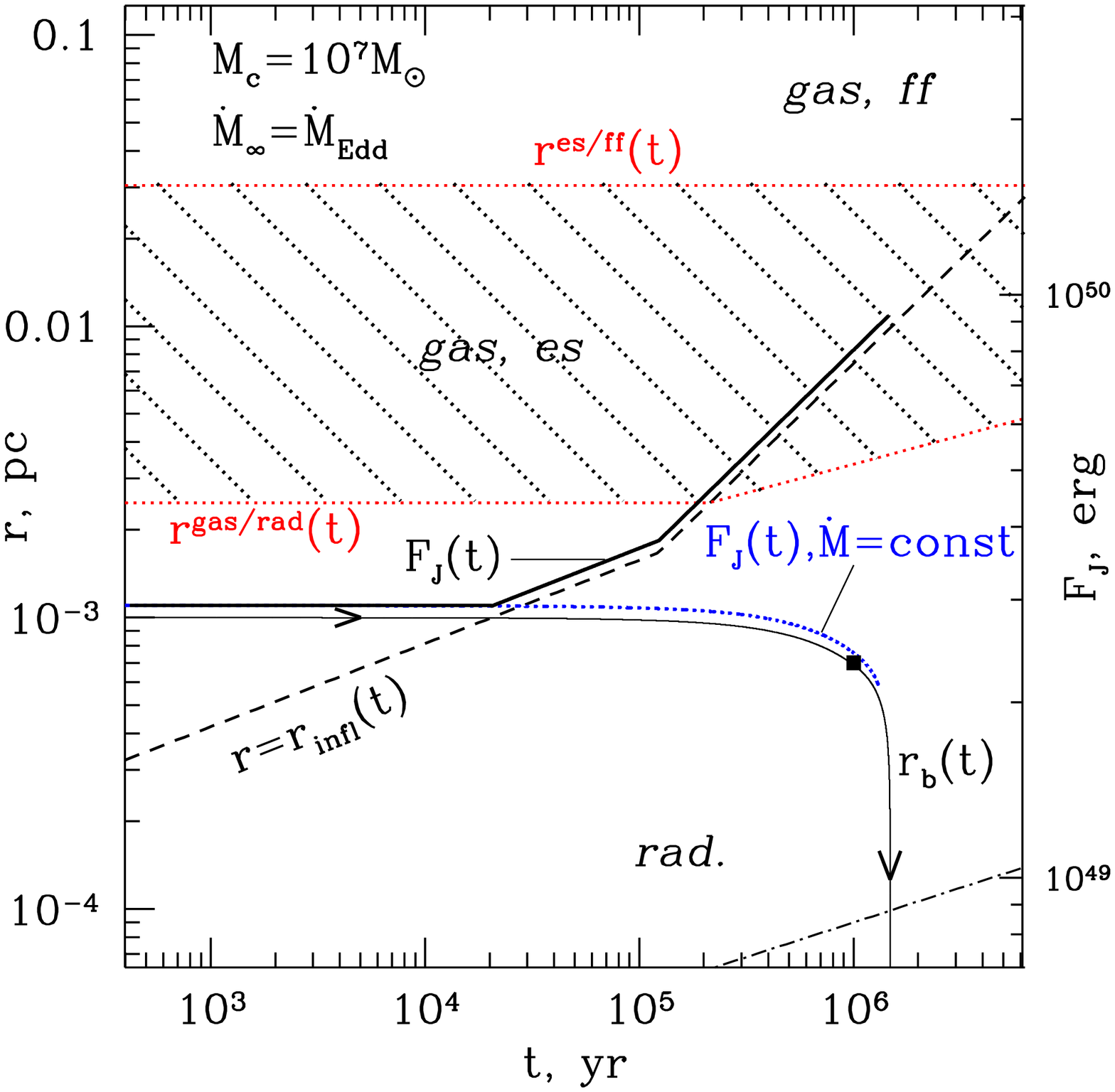}
\caption{
Characteristic evolution of the torque $F_J$ (labeled on the 
right axis) experienced by the central binary with $M_c=10^7$ 
M$_\odot$ and $q=1$ as a function 
of time. Two cases are shown: $F_J(t)$ computed self-consistently 
accounting for the circumbinary disk evolution (thick black solid line)
and assuming a standard constant $\dot M$ disk (blue dotted line).
In both cases the mass accretion rate in the disk far from the
binary is the same and equal to $\dot M_\infty=10^{-2}$ M$_{\rm Edd}$.  
Note the difference between the values of $F_J$ computed in two 
ways for $t\gtrsim 10^5$ yr.  
Also shown is the orbital evolution of the binary
(thin solid line) computed according to the first, self-consistent 
prescription for $\FJ(t)$. Different regimes in which the disk can 
be present are labeled on the plot (red dotted lines $r^{\rm rad/gas}$ 
and $r^{\rm es/ff}$ show their boundaries) and the gas pressure-dominated 
case with $\kappa=\kappa_{es}$ is shaded. Dashed curve shows 
the run of $r^{\rm infl}(t)$
in different regimes, and is closely related to the behavior of 
self-consistent $\FJ(t)$. Eddington limit becomes important below the
dot-dashed line in the lower right corner of the plot, see \S
\ref{sect:Edd_lim}. Black square 
dot marks the transition from the disk- to GW-dominated orbital 
evolution of the binary. See text for more details. 
\label{fig:typical}}
\end{figure}


\section{Implications for SMBH binary evolution.}  
\label{sect:SMBHs}

We now apply the results obtained in 
previous sections to the coupled evolution of SMBH binaries 
and disks around them. The two processes --- orbital
evolution of the binary and evolution of the disk properties 
--- must be considered simultaneously because of their  
mutual influence on each other. In exploring this evolution
one must pay special attention to the nonlocal effect
of the binary torque on the disk. 

Indeed, according
to the results obtained in \S \ref{sect:evol} the value 
of the angular momentum flux $F_J$ carried through the disk 
near the binary (which determines the orbital evolution 
of the binary) is not set locally but is determined by the 
disk properties at the radius $r_{\rm infl}$, which is the
outermost radius affected by the viscous transport of the 
angular momentum deposited in the disk by the binary. This 
property, often omitted in previous studies of the SMBH
binary evolution, is very important as we show further. 

We emphasize that the torque acting on the 
binary, which according to equation (\ref{eq:cond}) is 
equal to $\FJ(r_{in})$, is independent of the mass of the 
secondary $M_s$ and is the same irrespective of the 
mass ratio of the binary $q$ (see also IPP). Also, the 
dependence of $\FJ(r_{in})$ on the total mass of the 
binary $M_c$ arises only because $M_c$ determines the 
angular frequency in the disk. At first sight this may seem 
strange since the strength of tidal interaction is
determined by the potential of the secondary. However,
one has to keep in mind that the full torque exerted by 
the binary on the disk is generally found to scale as
(Goldreich \& Tremaine 1980; Papaloizou \& Lin 1984; 
Petrovich \& Rafikov 2012) 
\ba
\FJ(r_{in})\propto \frac{M_s^2\Sigma_0}{\Delta^3},
\label{eq:scaling}
\ea
irrespective of the precise form of the torque density
distribution. Here 
$\Sigma_0$ is the disk surface density just outside 
the region where the binary torques are important, and
$\Delta=|r_{in}-r_b|$ is the width of the gap --- the separation 
between the secondary orbit and the inner edge of the disk. 
Both $\FJ(r_{in})$ and $\Sigma_0$ are set globally at
the radius of influence. 

Equation (\ref{eq:scaling}) shows that a particular value 
of the binary 
torque $\FJ(r_{in})$ can be obtained not only by changing 
$M_s$ but also by varying the width of the gap $\Delta$ 
for a given $M_s$, and this is how the disk-binary tidal 
interaction self-regulates
itself to provide a necessary torque on the disk. As 
$\FJ(r_{in})$ varies in time for a fixed $M_s$ the width 
of the gap should also vary. The same is true if one varies
the mass of the secondary while keeping $\FJ(r_{in})$ 
constant --- the width of the gap would simply scale 
as $\Delta\propto q^{2/3}$. Of course, $q$ cannot be 
arbitrarily small since a low mass secondary may not be able 
to prevent the mass flow across its orbit (e.g. if 
$\Delta$ needed to provide a given value of $\FJ(r_{in})$ 
turns out being smaller than the disk scaleheight), 
meaning that the gap does not exist in the first place 
(see \S \ref{sect:overflow}). But as long
as the gap opening conditions are satisfied 
for a given $q$, the width of the
gap should always be able to adjust itself to provide just
the right amount of torque on the disk. 

In our subsequent calculations we will not be directly 
addressing the ``final pc'' problem (Lodato \etal 2009)
as we typically 
follow SMBH binaries starting at rather small separations,
$10^{-2}-10^{-4}$. Such binaries may be created by previous 
(possibly multiple) episodes of gas infall into the center 
of the galaxy in which the binary resides, each of which 
would tighten its orbit. At the same time some of our findings 
(e.g. significant reduction of the binary inspiral timescale
when the disk evolution is self-consistently included) 
are likely to be relevant for attempting to resolve 
the ``final pc'' problem by accounting for the possibility
of a circumbinary disk surrounding the binary. 

In all our calculations we take the viscosity
to scale with the total rather than the gas pressure in 
the radiation pressure dominated regime, i.e. $b=0$.
We will also assume that the tidal torque of the binary 
presents sufficiently strong barrier to inflowing gas 
to completely suppress gas overflow across the orbit of 
the secondary. This means that the boundary condition at 
the inner edge of the disk is given by 
$\chi=\dot M(r_{in})/\dot M_\infty=0$. In principle one 
can easily extend our results on the orbital evolution of
the binary to the case of $\chi\neq 0$.
Finally, even though the binary itself is not accreting 
when $\dot M(r_{in})=0$ the inner parts of the disk are 
still gaining mass, which changes the potential in 
which gas orbits further out. In this work we are mainly 
concerned with the disk-related effects on the binary 
evolution and for that reason we neglect the increase of 
the binary+disk system mass throughout the calculation.


\subsection{Binary inspiral: basic features.}  
\label{sect:inspiral}

We now look at the details of the orbital evolution of the 
binary SMBH embedded in a circumbinary disk. In Figure 
\ref{fig:typical} we show the joint variation
of the binary and the disk characteristics. The 
binary orbit is evolved according to 
\ba
\frac{dr_b}{dt}=-\frac{r_b}{t_{\rm GW}}-\frac{r_b}{t_J},
\label{eq:drbdt}
\ea
where $t_{\rm GW}$ and $t_J$ are given by equations
(\ref{eq:t_GW}) and (\ref{eq:t_J}). Variation of the 
disk properties, including the evolution of $\FJ(r_{in})$ 
entering the equation (\ref{eq:t_J}), is described below. 

We start 
an equal mass ($q=1$) binary with the total mass $M_c=10^7$ 
M$_\odot$ with initial semi-major axis of $10^{-3}$ pc. At 
time $t=0$ 
disk properties correspond to a standard
constant $\dot M=\dot M_{\rm Edd}$ disk extending 
from very large distances (effectively from infinity) down
to the binary semi-major axis (for simplicity we disregard 
the difference between the binary semi-major axis $r_b$ and
the inner radius of the disk, which is a factor of $2$ 
uncertainly at 
most, see MacFadyen \& Milosavljevi\'c 2008).

This setup would naturally arise if the
binary initially resided in a gas-free environment
and then gas started flowing into the galactic center  
in a disk-like configuration from large distances. Because 
viscous evolution 
accelerates at small radii the disk would naturally settle 
into a constant $\dot M$ configuration. At some point its 
inner radius would reach the vicinity of the binary and 
tidal torque would stop the gas inflow. This moment represents 
the starting point for our calculations. 

The radial dependence of disk properties at time $t=0$ can 
be found in Haiman \etal (2009) or by setting 
$F_J=\dot M_\infty(GM_c r)^{1/2}$ as appropriate 
for a standard constant $\dot M$ disk in formulae derived in 
\S \ref{sect:SMBH}. As Figure \ref{fig:typical} demonstrates
the binary starts in the radiation pressure dominated part of 
the disk but the transition to the gas pressure dominated 
regime occurs not too far outside of $r_b$, at 
$r^{\rm rad/gas}\approx 2.5\times 10^{-3}$ pc. 
Opacity switches from being dominated by the electron 
scattering to free-free opacity at $r^{\rm es/ff}\approx 
0.03$ pc. These regimes are clearly labeled in 
Figure \ref{fig:typical}. 

We also show the run of the radius of influence 
$r_{\rm infl}(t)$ in time by a dashed curve. At $r=r_{\rm infl}(t)$
the local viscous time equals the time since the start of 
the evolution $t$ but this is meaningful only if at time 
$t$ the disk extends to $r<r_{\rm infl}(t)$, which is
not always the case. Nevertheless, this dependence is 
still a useful concept as it allows us to see important 
transitions in the disk properties if we were to take 
an initial SMBH semi-major axis $r_b(0)$ different 
from the value shown in Figure \ref{fig:typical}.
The actual dependence $r_{\rm infl}(t)$ is calculated 
using definition (\ref{eq:l_infl}) and the expressions 
for $D_{J,0}$, $d$ and $n$ that can be found in 
Appendix \ref{sect:D_J}. Since these expressions are 
different in various physical regimes the behavior of
$r_{\rm infl}(t)$ exhibits distinct transitions as
it crosses the boundaries of different regimes, clearly 
visible in Figure \ref{fig:typical}.

\begin{figure*}
\centering
\includegraphics[width=17cm,height=9cm]{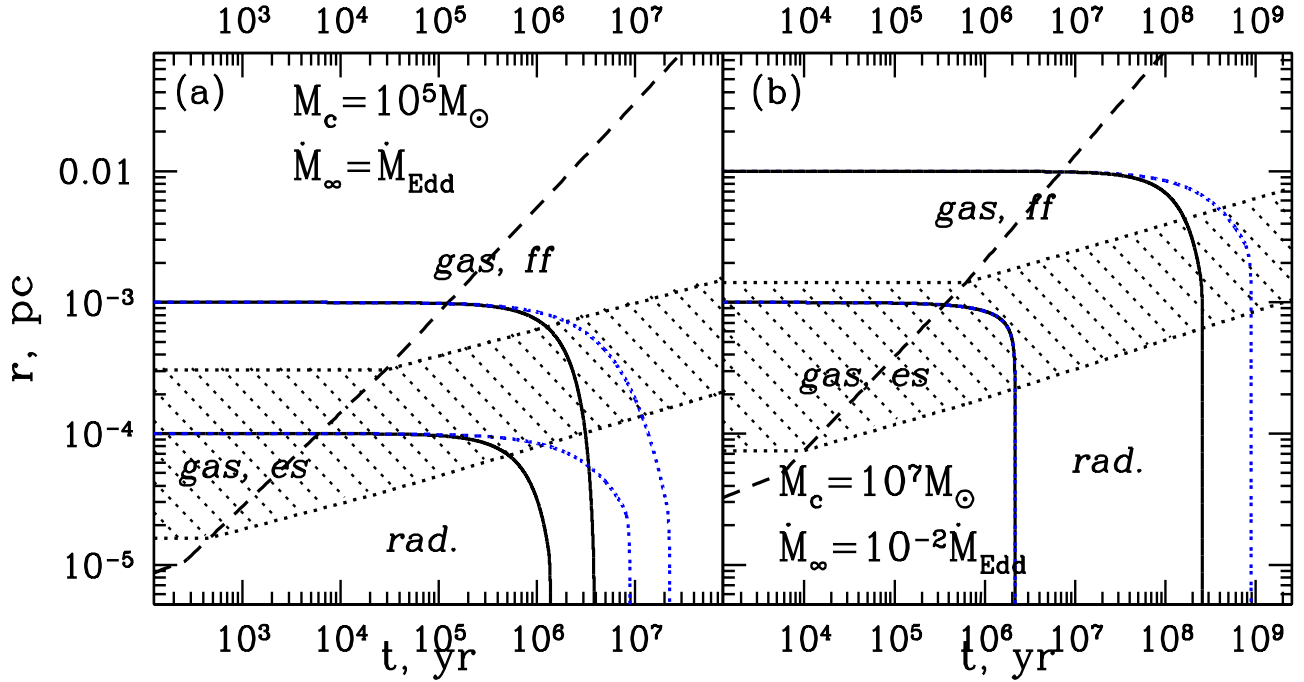}\\
\caption{
Comparison of the orbital evolution of the SMBH binary computed 
in two ways: fully accounting for the disk evolution driven by 
the tidal torque of the binary ({\it solid black curves}) and simply 
assuming the disk properties to be given by a standard constant 
$\dot M$ solution ({\it dotted blue curves}). In both cases the value
of $\dot M_\infty$, initial semi-major axis of the binary, and all other
parameters are assumed to be the same. Results are shown for 
$M_c=10^5$ M$_\odot$, $\dot M_\infty=\dot M_{\rm Edd}$ (panel a) and for
$M_c=10^7$ M$_\odot$, $\dot M_\infty=10^{-2}\dot M_{\rm Edd}$ (panel 
b), with binary mass ratio $q=1$ in all cases. Meaning of
other curves and labels on these plots is the same as in 
Figure \ref{fig:typical}.
Note that properly accounting for the binary-driven disk 
evolution can reduce the lifetime of the binary by almost an 
order of magnitude in some cases. 
\label{fig:comparison}}
\end{figure*}

As the binary starts tidally interacting with the disk the 
inflowing material begins accumulating near the inner edge 
of the disk. In addition to the evolution of the binary 
orbit (thin solid curve) this Figure also shows the time 
behavior of the torque $F_J(r_{in},t)$ at the inner edge of 
the disk (thick solid curve calibrated on the right axis), 
which is absorbed by the binary and causes its orbital 
evolution. The $F_J(r_{in},t)$ curve initially closely 
follows\footnote{For clarity we have slightly shifted the 
curve $F_J(r_{in},t)$ upward in Figure \ref{fig:typical} to avoid 
overlap with other curves, e.g. $r_b(t)$.} $r_b(t)$ 
because the torque exerted on the binary by a constant $\dot M$ 
disk is well approximated by $F_J(r_{in},t)=\dot M_\infty 
(G M_c r_b)^{1/2}$ as long as the inner edge of the disk 
tracks the binary orbit. This torque is small enough for $r_b$
not to change appreciably for rather long time. 

Approximately at $t=2\times 10^4$ yr, when $t_{visc}(r_b(0))\sim t$, 
the radius of influence of the binary torque $r_{\rm infl}$ 
grows beyond the initial
binary semi-major axis $r_b(0)$. As a result, angular momentum
flux at the inner edge of the disk $F_J(r_{in},t)$ starts increasing 
as $F_J(r_{in},t)\propto l_{\rm infl}(t) \propto t^{1/7}$
thus accelerating the orbital evolution of the binary. Unlike the 
calculations of the self-similar disk behavior in \S \ref{sect:self} 
in our present calculation we do not consider the details of the 
smooth transition between the parts of the disk inside and 
outside of $r_{\rm infl}$. Instead we simply assume that 
$F_J(r,t)= \dot M_\infty l$ for $l>l_{\rm infl}(t)$ as appropriate
for a constant $\dot M$ disk, while 
$F_J(r,t)= \dot M_\infty l_{\rm infl}(t)$ is constant in space 
for $l<l_{\rm infl}(t)$. In other words, we adopt a simple 
piece-wise dependence of $F_J$ on $r$:
\ba
F_J(r,t)= \left\{
\begin{array}{c}
\dot M_\infty
\left(G M_c r\right)^{1/2},~~~r>r_{\rm infl}(t),\\
\dot M_\infty
\left[G M_c r_{\rm infl}(t)\right]^{1/2},~~~r\le r_{\rm infl}(t),
\end{array}
\right.
\label{eq:piece}
\ea 
For that reason at late times $F_J(r_{in},t)$ starts tracking 
the run of $r_{\rm infl}(t)$ in Figure \ref{fig:typical}. 

Initially $r_{\rm infl}$ stays in the radiation pressure 
dominated regime and varies as 
$r_{\rm infl}\propto t^{2/7}$, see equations (\ref{eq:l_infl}) 
and (\ref{eq:DJ_b=0}). At $t\approx 2\times 10^5$ yr the radius 
of influence reaches out into the gas pressure dominated 
regime with $\kappa=\kappa_{es}$. There $r_{\rm infl}$ grows 
as $r_{\rm infl}\propto t^{5/7}$ (see the break in slope of 
$r_{\rm infl}$ curve), and the increase of $F_J(r_{in},t)$ 
and of the torque acting on the binary accelerates. 

\begin{figure*}
\centering
\includegraphics[width=17cm,height=16cm]{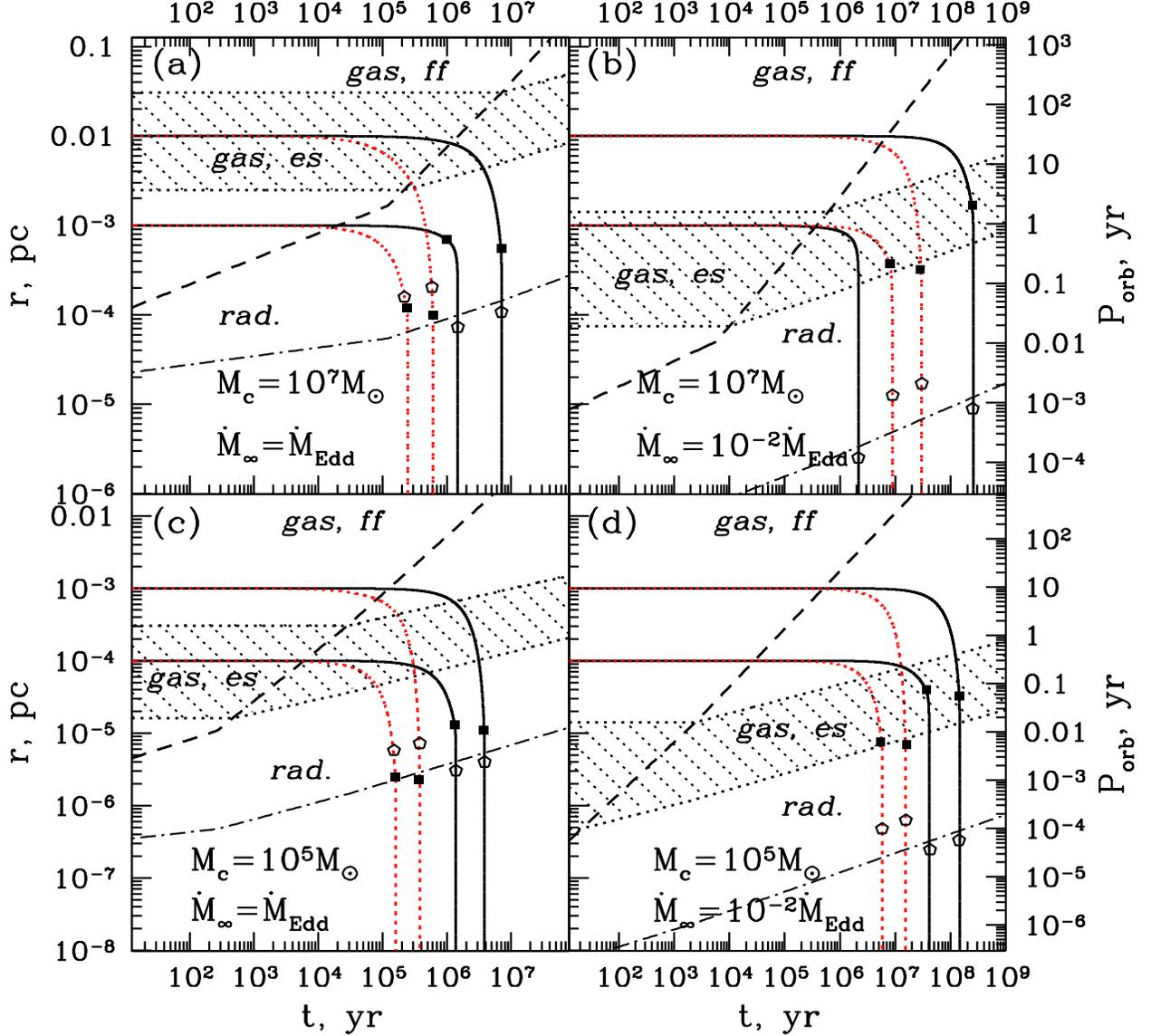}\\
\caption{
Evolutionary tracks of SMBH binaries for different values of $M_c=10^7$,
$10^5$ M$_\odot$ and $\dot M_\infty/\dot M_{\rm Edd}=1$, $10^{-2}$, 
labeled on panels. Tracks for $q=1$ ({\it solid black}) and $10^{-2}$ 
({\it dotted red}) are shown and in each panel we consider two 
starting values of the binary semi-major axis $r_b(0)$. Layout 
of these plots and meaning of different curves and labels are 
the same as in Figure \ref{fig:typical}. In particular, dashed
and dot-dashed curves show the run of $r_{\rm infl}$ (equation 
[\ref{eq:l_infl}]) and $r_{\rm Edd}$ (equation [\ref{eq:Edd_radius}]) 
correspondingly, and black square dots mark the transition
to the GW-dominated orbital decay. Open pentagons mark the
possible onset of the gas overflow, see \S \ref{sect:overflow}. 
See text for detailed description of evolution.
\label{fig:binary_evolve}}
\end{figure*}

In our calculations the disk interior to $r=r_{\rm infl}$ 
is a constant $F_J$ disk (see equation [\ref{eq:piece}]) 
with properties explored in  
\S \ref{sect:SMBH} rather than a constant $\dot M$ disk 
present outside of the radius of influence. As the value of $F_J$
in the inner disk grows the disk properties keep changing
as well. In particular the boundaries of the different 
regimes, i.e. $r^{\rm rad/gas}$ and $r^{es/ff}$,  
expand as the time goes by. Thus, the state of the disk at 
the location of the binary may change 
not only because of the variation of the binary orbit, 
but also due to the disk evolution, see 
Figures \ref{fig:comparison} \& \ref{fig:binary_evolve}. 

Note, that even though in the calculation shown in Figure 
\ref{fig:typical} the binary is always in contact with 
the radiation pressure dominated region of the disk, beyond 
$t\approx 2\times 10^5$ yr the torque 
on the binary is determined by the gas pressure dominated part 
of the disk with $\kappa=\kappa_{es}$, since this is where 
$r_{\rm infl}$ is. This demonstrates the 
nonlocality of the disk influence on the binary --- the state of 
the disk near the binary is essentially irrelevant for 
its orbital evolution. It is what goes on in the disk at 
$r\sim r_{\rm infl}$ that determines the torque acting 
on the SMBH binary at late times. 

This fact has been overlooked in previous studies of the 
gas-assisted SMBH inspiral problem. In particular, Haiman \etal 
(2009) and Kocsis \etal (2011) used the self-similar 
results of SC95 and IPP to account for the 
mass accumulation in the disk near the binary orbit. However,
in their calculations they have effectively assumed that
the radius of influence $r_{\rm infl}$ corresponds to the
same regime of the disk in which the binary is currently residing. 
In the case shown in Figure \ref{fig:typical} this would mean 
that even after $2\times 10^5$ yr the increase of $r_{\rm infl}$ 
with time would be calculated according to the scaling for the 
radiation pressure dominated regime, leading to an 
{\it underestimate} of the torque acting on the binary and an 
{\it overestimate} of its inspiral time. Our calculations fully
take into account the nonlocality of the disk influence on 
the disk.  

To better illustrate the role of mass puleup and disk evolution 
for the orbital evolution of the binary we compare our results
with calculations in which disk properties remain well represented
by the properties of a constant $\dot M$ disk at all times. Then
the torque on the binary is always given by $F_J(t)=\dot M_\infty 
\left[G M_c r_b(t)\right]^{1/2}$, see dotted curve 
in Figure \ref{fig:typical}. Clearly, this torque can be much 
smaller than the real $F_J(t)$, especially at late times,  
meaning that such calculations should overestimate the 
inspiral time of the binary. 

In Figure \ref{fig:comparison}
we display $r_b(t)$ calculated using constant $\dot M$ disk 
properties (dotted curves) and fully accounting for the 
binary-driven disk evolution (solid curves) for two equal 
mass ($q=1$) SMBH binary+disk systems: 
one with $M_c=10^5$ M$_\odot$, $\dot M_\infty=\dot M_{\rm Edd}$ 
and another with $M_c=10^7$ M$_\odot$,  
$\dot M_\infty=10^{-2}\dot M_{\rm Edd}$. Two different starting radii
$r_b(0)$ are explored in both cases. 

One clearly sees that in most cases evolution of $r_b$ computed
using standard constant $\dot M$ disk properties 
considerably overestimates (by almost an order of magnitude in some cases)
the binary inspiral time. In the case of $M_c=10^7$ M$_\odot$, 
$\dot M_\infty=10^{-2}\dot M_{\rm Edd}$, and $r_b(0)=10^{-3}$
pc the evolutionary tracks of the binary computed by two 
methods coincide. This is because in this case GW emission is 
more important for the evolution of the binary than the tidal 
coupling to the disk at all times, resulting in a universal 
behavior of $r_b(t)$. But whenever disk torques are important,
binaries shrink faster when the self-consistent disk evolution
is properly taken into account. For that reason, we strongly 
discourage the use of standard constant $\dot M$ disk properties 
for exploring the evolution of the central binary of a 
circumbinary disk.

\subsection{Binary inspiral: parameter exploration.}  
\label{sect:inspiral_pars}

\begin{figure*}
\centering
\includegraphics[width=17cm,height=18cm]{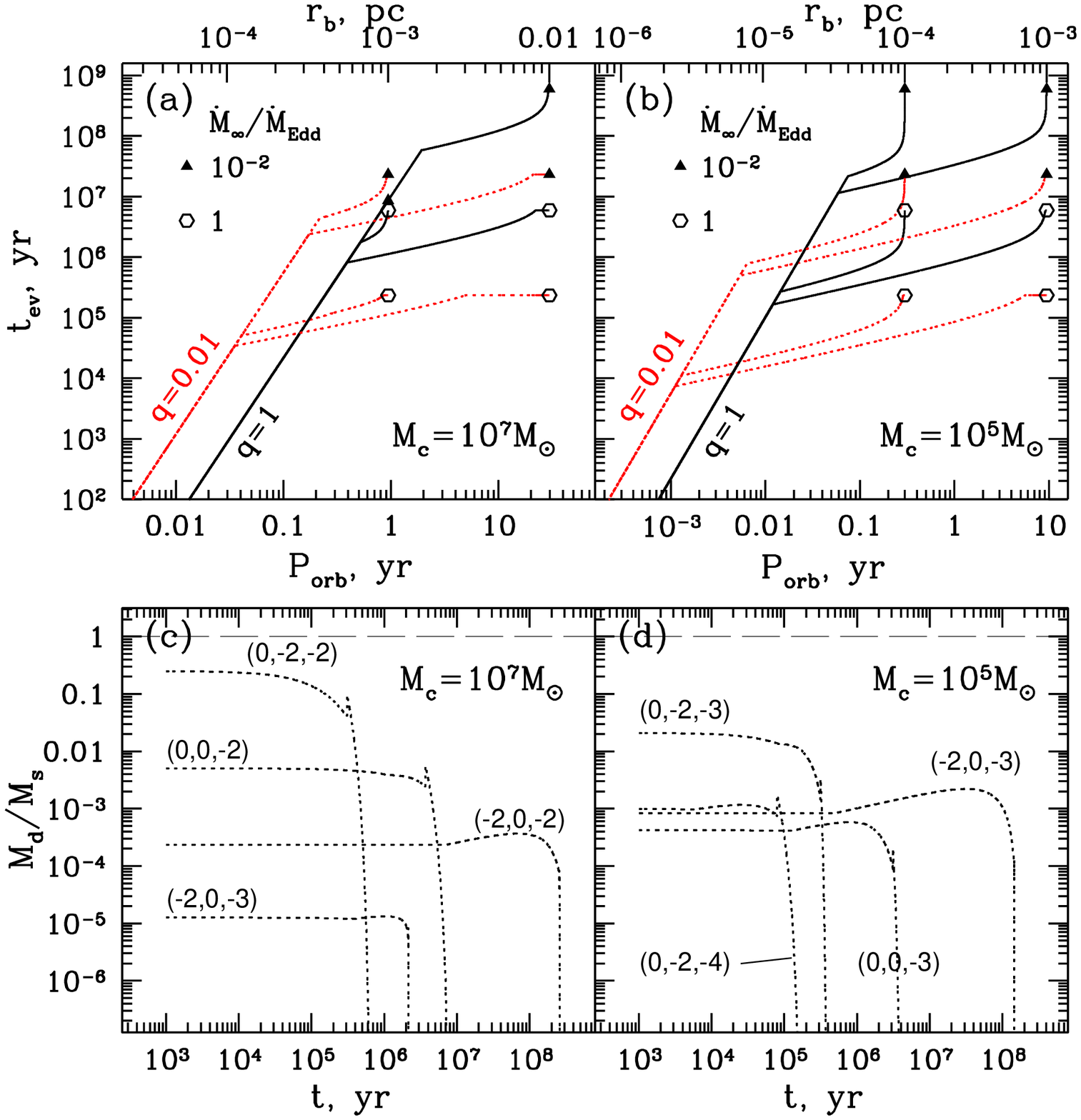}\\
\caption{
(a,b) Evolution timescale $t_{ev}=|d\ln r_b/dt|^{-1}$ of SMBH 
binaries as a function of orbital period $P_{orb}$ or semi-major 
axis $r_b$ (upper axis; use to read off the initial semi-major axis 
$r_b(0)$). Starting points for each evolutionary track are marked 
by either triangles (for $\dot M_\infty=10^{-2}\dot M_{\rm Edd}$) 
or hexagons (for $\dot M_\infty=\dot M_{\rm Edd}$). Solid (black) and 
dotted (red) curves are used for $q=1$ and $q=10^{-2}$ binary tracks.
Straight portions of these tracks at low $P_{orb}$ correspond to
the epoch when the orbital evolution of the binary is dominated 
by the GW emission. Note that prior to that tracks starting at 
different $r_b(0)$ do not overlap, illustrating hysteresis of
the binary evolution caused by the non-local nature of the 
self-consistent disk-binary coupling.  
(c,d) Ratio of the local disk mass $M_d$ to the secondary 
mass $M_s$ for a sub-sample of evolutionary tracks displayed in
Figure \ref{fig:binary_evolve}, 
showing that $M_d/M_s\ll 1$ in our calculations. 
Different tracks are labeled as
$\left(\log[\dot M_\infty/\dot M_{\rm Edd}],\log q,
\log[r_b(0)/{\rm pc}]\right)$. See text for details.
\label{fig:time_evolve}}
\end{figure*}

In Figures \ref{fig:binary_evolve} \& \ref{fig:time_evolve}
we provide a more detailed and systematic view of the SMBH 
binary evolution under different conditions. 
We explore two representative values of $M_c$: $10^5$ M$_\odot$ 
(implying Schwarzschild radius $R_S=10^{-8}$ pc) and 
$10^7$ M$_\odot$ ($R_S=10^{-6}$ pc), but our results can 
be trivially extended to other values of $M_c$.
Accretion rate in the disk at large distances is taken to 
be either $\dot M_\infty=10^{-2}M_{\rm Edd}$ or $M_{\rm Edd}$, 
and binary mass ratio is varied between $q=10^{-2}$ and $1$. 
We also consider two different values of 
the starting semi-major axis of the binary $r_b(0)$: $10^{-4}$
pc and $10^{-3}$ pc for $M_c=10^5$ M$_\odot$, and $10^{-3}$
pc and $10^{-2}$ pc for $M_c=10^7$ M$_\odot$. These values are 
close to the ``bottleneck'' semi-major axes at which the stellar 
dynamical orbital evolution of SMBH pairs decelerates 
dramatically, see Yu (2002).

One might worry that the outer parts of the disk can be prone to
gravitational instability. We determined that in the initial 
constant $\dot M$ disk Toomre $Q$ equals unity at 
$r^{sg}=9\times 10^{-3}$ pc, $0.3$ pc, $0.3$ pc, and $0.5$ pc
for the systems shown in panels (a)-(d) correspondingly in 
Figure \ref{fig:binary_evolve} (for $\alpha=0.1$, 
$\varepsilon=0.1$, $\mu=0.5m_p$). Thus, at the start of our 
calculations one needs to worry about the importance of the 
disk self-gravity only for the system with $M_c=10^7$ M$_\odot$, 
$\dot M_\infty=M_{\rm Edd}$ and the binary starting at 
$r_b(0)=10^{-2}$ pc, where the disk can be marginally gravitationally 
unstable. One has to keep in mind though that later the torque on
the binary is going to be set at the radius of influence that 
expands beyond $r_b(0)$, see \S \ref{sect:inspiral}. If 
$r_{\rm infl}$ would exceed $r^{sg}$ at some point, the 
calculations of disk evolution would need to be refined. In 
our present study we neglect this complication; it may only 
be an issue for the high-mass systems shown in 
Figure \ref{fig:binary_evolve}a. 

Figure \ref{fig:binary_evolve} shows evolutionary tracks of 
the binary orbit mapped onto the disk state in the format 
analogous to that used in Figures \ref{fig:typical} \& 
\ref{fig:comparison}. In addition, in Figure \ref{fig:time_evolve}a,b
we show the dependence of the binary orbital evolution  
timescale $t_{ev}\equiv -r_b/\dot r_b$ vs. the binary orbital
period $P_{orb}$ or semi-major axis $r_b$ (upper axis) for 
the evolutionary tracks displayed in Figure \ref{fig:binary_evolve}. 
This plot allows us to easily see the transition from the disk 
dominated evolution at longer periods to the GW dominated phase, 
which is clearly described by straight line tracks at small
values of $P_{orb}$. Figure \ref{fig:time_evolve}a,b can be 
directly compared to the analogous 
$t_{ev}(P_{orb})$ plots in Haiman \etal (2009). 

Results presented in Figures \ref{fig:binary_evolve} \& 
\ref{fig:time_evolve} can be summarized in the following 
set of conclusions.


\subsubsection{Circumbinary disk can be efficient in driving 
orbital evolution of the binary.}  
\label{sect:disk_efficient}

We generally agree with the results of existing studies
(e.g. IPP, Haiman \etal 2009) that disks can appreciably 
accelerate orbital evolution of SMBH binaries. Figures
\ref{fig:binary_evolve} \& \ref{fig:time_evolve} clearly 
illustrate this point. According to equation (\ref{eq:t_GW}) 
without the disk the orbital evolution timescale due to 
emission of gravitational waves is rather long for some of 
the systems shown in these plots: for example, 
$t_{\rm GW}=8.3\times 10^{10}$ yr for equal mass $10^7$ M$_\odot$
SMBH binary starting at $r_b(0)=10^{-2}$ pc. Systems with small 
mass ratios evolve even slower: a binary with $q=10^{-2}$  
takes $\approx 25$ times longer to merge due to the GW 
emission alone. The only system in our sample strongly 
affected by the GW emission from the very start is the 
equal mass $10^7$ M$_\odot$ binary starting at $10^{-3}$ pc 
and surrounded by a disk with $\dot M_\infty=10^{-2}\dot 
M_{\rm Edd}$, see Figure \ref{fig:time_evolve}a. 
As expected this system merges faster than its $q=10^{-2}$ 
counterpart, even though the latter is affected more by the 
disk torques. 

On the other hand, the same $q=1$, $10^7$ M$_\odot$ binary 
starting at $r_b(0)=10^{-2}$ pc and surrounded a disk 
accreting at $\dot M_\infty=\dot M_{\rm Edd}$ merges within 
$7$ Myr --- more than 4 orders of magnitude faster than 
without the disk! Lower mass disks are of course less 
efficient at driving orbital evolution of SMBH binaries --- 
the same binary surrounded 
by a disk with $\dot M_\infty=10^{-2}\dot M_{\rm Edd}$ merges
within $3\times 10^8$ yr, but this is still much shorter than
the corresponding $t_{\rm GW}$. Lower mass ratio binaries are 
affected by the disk even stronger, provided that they present
an efficient barrier to the mass inflow at the orbit of the 
secondary. For example, $10^7$ M$_\odot$ binary with $q=10^{-2}$
surrounded by a disk with $\dot M_\infty=10^{-2}\dot M_{\rm Edd}$, 
merges within $3\times 10^7$ yr, about an order of magnitude 
faster than the $q=1$ binary with the same parameters.

In Figure \ref{fig:binary_evolve} black square dots mark the 
location on the evolutionary track of each SMBH binary 
where the disk dominated evolution switches to the GW dominated
orbital decay. In systems with massive, high-$\dot M_\infty$
disks this transition typically occurs when the binary is
in contact with the radiation pressure dominated part of the 
disk. However, in systems with less massive, lower-$\dot M_\infty$
disks this transition may happen even while the binary is
surrounded by the gas pressure dominated (with $\kappa=\kappa_{es}$)
part of the disk. 

Another way to state the importance of the 
disk dominated evolution is to note that in most cases
transition to GW dominated regime occurs at $r_b\ll r_b(0)$,
i.e. after the binary semi-major axis has been significantly
reduced by the disk torques. For example, evolution of a 
$M_c=10^5$ M$_\odot$ binary with $q=10^{-2}$ starting at 
$10^{-3}$ pc is dominated by torques produced by a 
$\dot M_\infty=\dot M_{\rm Edd}$ disk down to 
$r_b\approx 2\times 10^{-6}$ pc, see Figure
\ref{fig:binary_evolve}c. This is almost three orders of 
magnitude smaller than $r_b(0)$ and is about 200 $R_S$ for
the binary. This additionally emphasizes the important role 
of disk torques in shrinking the SMBH binary orbits, even 
at relatively large separations.


\subsubsection{Nonlocal character of the disk-binary coupling}
\label{sect:nonlocal}

As described in \S \ref{sect:inspiral} the torque exerted on the 
binary is set by the disk properties at 
$r_{\rm infl}$. Initially binary can affect only its immediate 
surroundings as it takes certain time for the disk to absorb 
enough angular momentum injected by the binary to affect 
the surface density distribution further out in the disk.
For that reason initially $F_J(r_{in})$ is essentially the 
same as in the case of a constant $\dot M=\dot M_\infty$ disk, 
and is set by the disk {\it locally}, at 
$r\sim r_b$. This torque is usually rather small implying
slow orbital evolution and long values of $t_{ev}$.

However, after system has evolved for time comparable to the 
viscous timescale at the inner disk edge $t_{visc}(r_{in})$, 
$r_{\rm infl}$ starts exceeding $r_b$. Past that point the 
torque on the binary $F_J(r_{in})$ is being set {\it globally}, 
at distances far exceeding $r_b$, which can be clearly seen in 
several evolutionary tracks shown in Figure 
\ref{fig:binary_evolve}. Increase of $F_J(r_{in})$ initially 
occurs at almost constant $r_b$, which is reflected in almost
vertical initial evolutionary tracks in $t_{ev}-P_{orb}$ plane
shown in Figure \ref{fig:time_evolve}a,b, clearly 
visible for $M_c=10^5$ M$_\odot$ and low 
$\dot M_\infty/\dot M_{\rm Edd}=10^{-2}$. 

Global nature of the torques is generally 
more pronounced for lower mass binaries and for lower 
$\dot M_\infty$. This is because higher $M_c$ implies earlier 
transition to the GW-dominated orbital decay (see Figure 
\ref{fig:time_evolve}a,b), shortening the binary lifetime and 
not allowing $r_{\rm infl}$ to extend as far as in the 
lower $M_c$ case. Higher $\dot M_\infty/\dot M_{\rm Edd}$ 
plays similar role, shortening the binary lifetime and 
reducing the radius of influence at the end of inspiral 
compared to the lower $\dot M_\infty$ case. This is true even 
though $r_{\rm infl}$ itself depends on both $M_c$
and $\dot M_\infty$ --- these dependencies are usually rather 
weak, see Appendix \ref{sect:D_J}.

Almost all tracks shown in Figure \ref{fig:binary_evolve} 
at some point in their evolution run into the situation 
described in \S \ref{sect:inspiral}, where 
the inner edge of the disk (assumed equal to $r_b$) and 
$r_{\rm infl}$ reside in parts of the disk corresponding to 
different physical regimes. This means that calculations
of the binary evolution assuming the scaling of 
$r_{\rm infl}(t)$ to always correspond to the physical state
of the disk near the binary orbit (Haiman \etal 2009; Kocsis 
\etal 2011) are not accurate. Figure \ref{fig:binary_evolve} 
demonstrates that using this simple-minded procedure for a 
binary in the radiation pressure-dominated regime can
easily underestimate the torque $\FJ(r_{in})$ driving its 
orbital evolution thus overestimating the binary lifetime
(see e.g. Figure \ref{fig:binary_evolve}a,c). 
Analogously, for a binary residing in the gas pressure 
dominated regime with $\kappa=\kappa_{es}$ and $r_{\rm infl}$
extending into the part of the disk where $\kappa=\kappa_{ff}$
(see Figure \ref{fig:binary_evolve}b-d) the use of 
$r_{\rm infl}(t)$ scaling corresponding to the binary location 
would again lead to a (mild) underestimate of  $\FJ(r_{in})$ 
($r_{\rm infl}(t)\propto t^{4/5}$ when $\kappa=\kappa_{ff}$) 
and an overestimate of the merger time.

Another consequence of the non-locality of the disk-binary 
coupling is the clear {\it hysteresis} in the evolution of the 
system --- the dependence of the current rate of orbital decay 
of the binary on the previous history of the disk evolution. 
This property is most readily seen in Figure 
\ref{fig:time_evolve}, in which the evolutionary tracks computed 
for the same $M_c$, $q$, $\dot M_\infty$ but starting at 
different initial radii do not result in the same orbital 
decay timescale $t_{ev}$ at a given orbital period in the 
disk-dominated regime. One can see that for a fixed $P_{orb}$ 
the inspiral timescales computed for different starting 
conditions can differ by a factor of several. This is in contrast
to local calculations presented in Haiman \etal (2009), in which
orbital evolution depends only on the current value of $r_b$, 
see their Figures 1-5 showing just a single 
evolutionary track for a given set of $M_c$ and $q$.


\subsubsection{Evolution of the disk accelerates orbital 
evolution of the binary}
\label{sect:accelerate}

Because the radius $r_{\rm infl}$ setting the value of the 
inner torque $\FJ(r_{in})$ in our calculations (see 
equation [\ref{eq:piece}]) steadily grows, we necessarily 
find the disk evolution to result in the {\it speed up} of 
the binary decay compared to the case in which the disk 
is a constant $\dot M=\dot M_\infty$ disk at all times. 
As we have shown in \S \ref{sect:inspiral} this results in 
shorter lifetime of the system in our calculations,
which is a very natural result.
 
Interestingly, previously Haiman \etal (2009) have reached 
a directly opposite conclusion --- that the disk evolution 
caused by the mass 
pile up near the orbit of the secondary {\it slows down}
its inspiral (see their Figures 6 \& 7). 
Whether this difference is caused by the local character 
of the of the disk-binary coupling assumed by Haiman 
\etal (2009) or by the adoption of SC95 
solution in their calculations is not clear.


\subsubsection{Validity of the secondary dominated regime}
\label{sect:secondary}

In Figure \ref{fig:time_evolve}c,d we display the ratio 
of the secondary mass $M_s$ to the
local disk mass $M_d=\Sigma(r_b)r_b^2$. Our calculations
explicitly assume a ``secondary-dominated'' limit 
$M_d/M_s\lesssim 1$ (SC95; IPP; Haiman \etal 
2009) since only in this case
the boundary condition in the form (\ref{eq:cond}) is valid, 
as discussed in \S \ref{sect:BCs}. 
One can easily see that almost 
all evolutionary tracks shown in Figure 
\ref{fig:time_evolve}c,d (which are also present 
in Figure \ref{fig:binary_evolve}) satisfy $M_d/M_s\ll 1$ and 
thus correspond to the secondary-dominated regime. This finding 
is in agreement with Haiman \etal (2009) conclusion on the 
ubiquity and importance of this stage of the binary evolution. 

Evolutionary track for a $q=10^{-2}$, $M_c=10^7$ M$_\odot$ 
SMBH binary starting at $10^{-2}$ pc and surrounded by a disk 
accreting at $\dot M_\infty=\dot M_{\rm Edd}$ shows the highest 
$M_d/M_s\sim 0.2$ at early stages of the evolution among all
the tracks computed in this work. For this binary the 
secondary-dominated regime is close to being marginally
violated, and its track exhibits noticeable evolution
even prior to crossing $r_{\rm infl}(t)$ curve. 
Whenever $M_d\sim M_s$ the inner part 
of the disk contains the amount of angular momentum 
comparable to the full orbital angular momentum of 
the binary at the very start of evolution. Thus, to cause 
appreciable evolution of the disk via tidal coupling (which
happens on the local viscous timescale) and
to push the radius of influence beyond $r_b(0)$ requires the
binary to give off a noticeable fraction of its orbital 
angular momentum. As a result a massive disk is able
to shrink the binary orbit very efficiently, in less than 
the local viscous timescale at the inner edge of the disk. 
 
At even higher values of $M_d/M_s$ one needs
to employ a boundary condition (\ref{eq:intMdot}) different 
from (\ref{eq:cond}) to 
properly compute the disk evolution.


\subsubsection{Importance of the Eddington limit}
\label{sect:Edd_lim}

As we discussed in \S \ref{sect:F_J} the magnitude of the 
angular momentum flux in a constant $\FJ$ disk cannot be 
arbitrarily high, because of the existence of the Eddington 
limit (\ref{eq:FJ_Edd}) in the radiation pressure dominated 
region of the disk. We now check how important is this 
constraint for the SMBH binary evolution.

In Figure \ref{fig:binary_evolve} we plot as a dot-dashed 
curve the Eddington radius $r_{\rm Edd}$ at which the Eddington limit 
becomes important. This dependence can be easily derived 
from equation (\ref{eq:FJ_Edd}) and reads
\ba
r_{\rm Edd}(t)&=&\left[\frac{\kappa_{es} F_J(t)}
{2\pi c(GM_c)^{1/2}}\right]^{2/3}
\nonumber\\
&=&\left(\frac{\kappa_{es}\dot M_\infty}
{2\pi c}\right)^{2/3}r_{\rm infl}^{1/3}(t),
\label{eq:Edd_radius}
\ea
where the value of $\FJ$ in the inner part of the disk affected
by the binary torque is assumed to be given by equation 
(\ref{eq:piece}). In this expression we use the dependence 
$r_{\rm infl}(t)$, which is also plotted in the same Figure.
Eddington limit is important for a 
circumbinary disk whenever $r_b<r_{\rm Edd}(t)$. 

Figure \ref{fig:binary_evolve} demonstrates that for our 
choices of $M_c$, $q$, $\dot M_\infty$ and $r_b(0)$ the
Eddington limit is essentially irrelevant during the 
disk-dominated phase of the orbital evolution of the binary: 
dot-dashed curve of $r_{\rm Edd}(t)$ always passes
below or close to the square dots marking the transition to 
the GW-dominated regime. This provides justification for our
calculations of the binary evolution since the details of 
the disk physics, including the advent of the Eddington limit,
are not going to affect it in the GW-dominated regime.

On the other hand, for all evolutionary tracks depicted in 
Figure \ref{fig:binary_evolve} the Eddington limit does become 
important at some point and affects the disk properties 
right before the binary merger. This may have important
implications for the electromagnetic precursor of the 
merger. 

It is worth stressing that the Eddington limit 
in a constant $\FJ$ disk can be easily reached even if the
mass accretion rate in the disk at large separations is far
less than $\dot M_{\rm Edd}$, see Figure \ref{fig:binary_evolve}b,d. 
This is obviously caused by accumulation of mass at the 
inner edge of the disk resulting in higher temperature and 
more important 
role of the radiation pressure than in a constant $\dot M$
disk. Thus, the Eddington limit in self-consistently 
evolved circumbinary disks is more stringent than in the 
standard constant $\dot M$ disks.

At the same time, the Eddington limit is still
less important for low values of $\dot M_\infty/\dot M_{\rm Edd}$:
for $\dot M_\infty=10^{-2}\dot M_{\rm Edd}$
it kicks in only at the distances of several tens of Schwarzschild
radii at most, see \ref{fig:binary_evolve}b,d. At the same time for 
$\dot M_\infty=\dot M_{\rm Edd}$ the Eddington limit starts affecting 
the disk near the binary as soon as the latter enters the 
GW-dominated regime.


\subsubsection{Gas overflow across the orbit of the 
secondary}
\label{sect:overflow}

Our assumption\footnote{I am grateful to Zoltan Haiman for 
suggesting the overflow calculations presented in this section.} 
of $\dot M(r_{in})=0$ adopted throughout most of 
this work is equivalent to demanding the width of the 
gap between the orbit of the secondary and the inner edge of 
the disk $\Delta$ to be larger than the disk scaleheight $h$. 
Indeed, the torque density produced by the planet drops for
$|r-r_b|\lesssim h$ due to the phenomenon of the
``torque cutoff'' (Goldreich \& Tremaine 1980), which implies
that the secondary can effectively repel the disk
fluid only if the gap width satisfies $\Delta\gtrsim h$. When 
this condition is not 
fulfilled, gas enters the torque cutoff zone near the orbit of 
the secondary where the tidal repulsion is no longer effective
and starts to overflow the orbit of the secondary 
(Kocsis \etal 2012). As a result, the inner boundary condition 
in the form $\dot M(r_{in})\ll \dot M_\infty$ may get violated. 

Following SC95 we can estimate the gap width
as
\ba
\frac{\Delta}{h}\sim\left[\frac{q^2}{\alpha\beta^b}
\left(\frac{r}{h}\right)^5\right]^{1/3},
\label{eq:Delta_h}
\ea
where we assumed viscosity to be given by equation 
(\ref{eq:nu}). In both the gas 
pressure and the radiation pressure dominated 
regime with $b=0$ (which we adopt in our calculations) one obtains
the same value of $\Delta$ as in SC95. However, 
in the radiation pressure dominated regime with $b=1$ (as adopted 
e.g. by Kocsis \etal 2012a,b) the gap is wider by a factor of
$\beta^{-1/3}$, making overflow less likely. 

Liu \& Shapiro (2010) calculated mass accretion rate across 
the orbit of the secondary using a local steady state model for
the disk structure. They found that 
$\dot M(r_{in})$ is exponentially small when the 
factor\footnote{Liu \& Shapiro (2010) call this factor $\tilde g$.}
in square brackets in equation (\ref{eq:Delta_h}) is large 
($\gtrsim 5-10$). On the contrary, when this factor is 
$\lesssim 5$ the mass accretion rate across the orbit of 
the secondary is found to be close to $\dot M_\infty$.
This provides justification for our use of the boundary condition 
$\dot M(r_{in})=0$ ($\chi=0$) whenever $\Delta/h\gtrsim 1$, and 
for assuming overflow to occur for $\Delta/h\lesssim 1$.

The condition (\ref{eq:Delta_h}) allows us to find the value 
of $r$ at which $\Delta/h=1$, which we call the ``overflow'' 
radius $r_{of}$.
Assuming that when the overflow begins the disk around the 
binary is in the radiation pressure dominated regime (this is 
always the case in our calculations, see Figure  
\ref{fig:binary_evolve}) with $b=0$ and $h/r$ is given by equation 
(\ref{eq:h_rad_es}), we find that $\Delta/h=1$ at
\ba
r_{of}\approx r_{\rm Edd}
\left(\frac{\alpha}{q^2}\right)^{2/15}, 
\label{eq:r_of}
\ea
where $r_{\rm Edd}$ is given by equation (\ref{eq:Edd_radius}).

It is clear from this expression that for equal mass binaries
overflow occurs only {\it after} the Eddington limit becomes
important since then $r_{of}\lesssim r_{\rm Edd}$. On the other 
hand,  as long as $q\lesssim \alpha^{1/2}$ one finds 
$r_{of}\gtrsim r_{\rm Edd}$
but the actual value of $r_{of}$ never deviates too much (i.e.
not by orders of magnitude) from 
$r_{\rm Edd}$ because of the weak dependence of $r_{of}$ on 
$q$ and $\alpha$. Indeed, for $\alpha=0.1$ and $q=10^{-2}$
one finds $r_{of}\approx 2.5 r_{\rm Edd}$.

In Figure \ref{fig:binary_evolve} we show the position of 
$r_{of}$ for each evolutionary track with open pentagons. 
These tags always lie close
to the dot-dashed curve showing the Eddington limit 
(\ref{eq:Edd_radius}). Another important observation 
that can be made by inspecting this Figure is that 
at least for $q>10^{-2}$ overflow always occurs close to the
start of the GW-dominated phase of the orbital evolution \
of the binary. Indeed, only for $q=10^{-2}$ and only in massive 
disks with $\dot M_\infty=\dot M_{\rm Edd}$ do we find that 
overflow {\it precedes} (by only a factor of $\sim 2$ in 
terms of $r_b$) the stage of the GW-driven evolution, see Figure 
\ref{fig:binary_evolve}a,c. Both for $q\sim 1$ and for 
$\dot M_\infty\ll \dot M_{\rm Edd}$ (essentially irrespective 
of $q$) overflow occurs when the orbital evolution
of the binary is already fully determined by the GW emission.

Kocsis \etal (2012) have derived quasi-steady solutions for 
the disk structure in presence of the overflow, reminiscent of 
the SC95 results. For these solutions to 
become valid after the overflow begins, information on the 
new boundary condition at the inner edge of the disk must 
propagate to the current radius of influence $r_{\rm infl}$ 
where the unperturbed, standard constant $\dot M$ disk starts.
Otherwise the solution would not converge to a standard constant 
$\dot M=\dot M_\infty$ solution at large radii and $\dot M$ 
at the inner edge of the disk cannot be assumed equal to 
$\dot M_\infty$. Establishing connection to the outer 
disk takes of order the viscous time at $r_{\rm infl}$, which
is about the age of the system in our calculations. Using Figures
\ref{fig:binary_evolve} and \ref{fig:time_evolve} one can 
easily see that unless $q\lesssim 10^{-2}$ and 
$\dot M_\infty=\dot M_{\rm Edd}$ the orbit of the binary
evolves on a much shorter timescale than the viscous time at 
$r_{\rm infl}$. 

For example, a disk with $\dot M_\infty=\dot M_{\rm Edd}$ 
around a binary with $M_c=10^7$ M$_\odot$,
$q=10^{-2}$ starting at $0.01$ pc begins to
overflow when its period is $\approx 0.1$ yr (see Figure 
\ref{fig:binary_evolve}a) and the orbital evolution time scale 
is $\approx 7\times 10^5$ yr (Figure \ref{fig:time_evolve}a).
This is almost the same as the viscous time at $r_{\rm infl}$
for the corresponding evolutionary track. Thus, a quasi-steady 
solution can be marginally valid in this case. But if we now 
look at $q=1$ binary keeping everything else the same we find 
overflow to occur at $P_{orb}=0.03$ yr. At this period  
$t_{ev}\approx 10^3$ yr which is much shorter than the viscous
time at $r_{\rm infl}$ ($\sim 10^7$ yr). As a result, a global 
quasi-steady solution does not get established in this case. 
Similar situation occurs for $\dot M_\infty\ll\dot M_{\rm Edd}$
(and arbitrary $q$).  

To summarize, the overflow across the orbit of the secondary 
is most important for low $q$, high $\dot M_\infty$ systems.
Even then it does not strongly affect the orbital evolution of 
the binary during the disk-driven stage and becomes 
truly important only when the binary inspiral is dominated 
by the GW emission. These conclusions are reached in a setup
most favorable for the emergence of the overflow --- radiation
pressure dominated disk with $\nu$ proportional to the radiation 
pressure $p_r$. If instead $\nu$ scales with gas pressure the 
overflow is going to be even less important for the gas-assisted
SMBH binary evolution.


\subsubsection{Implications for the gravitational wave 
signatures of SMBH binary merger.}
\label{sect:GW_imply}

Our results have interesting implications for future 
space-based gravitational wave antennae such as LISA.
If we adopt $0.03$ mHz ($\approx 10^{-3}$ yr$^{-1}$) as 
a characteristic lowest frequency probed by such 
experiments then according to Figure 
\ref{fig:time_evolve}a,b equal mass binaries with 
$M_c\gtrsim 10^4$ M$_\odot$ are detectable only
when their orbital decay is already fully dominated 
by the GW emission, even for massive disks 
with $\dot M_\infty=\dot M_{\rm Edd}$. 

However, lower mass ratio systems including the so-called
extreme mass ratio inspirals (EMRIs) can enter the detection 
band of the GW experiments during the stage when their orbital 
evolution is still dominated by the tidal coupling to the disk. 
An example of this can be seen in Figure \ref{fig:time_evolve}b
where the SMBH binaries with $q=10^{-2}$ are pushed
by the disk with $\dot M_\infty=\dot M_{\rm Edd}$
all the way until the transition to the GW-dominated regime 
occurs at $P_{orb}\approx 10^{-3}$ yr. Decay of systems with 
even lower $q$ or $M_c$ can be dominated by their disks down
to even shorter orbital periods, making detection of disk-driven
migration quite plausible for low $M_c$ EMRIs (provided that 
the strain they produce is above the signal-to-noise of the GW 
antenna).

Disk effects on the GW signal manifest 
themselves via the orbital phase shift of the binary caused
by the variation of its semi-major axis due to the disk-driven 
migration, see Kocsis \etal (2011), Yunes \etal (2011). 
Even if signatures of the disk-driven migration are indeed
found in the GW signal of coalescing binaries, it is unlikely
that one would be able to use these measurements to probe the 
properties of the radiation pressure dominated part of the 
disk in the immediate vicinity of the binary. The reason for 
that again lies in the nonlocal nature of the torque acting 
on the binary: close to $P_{orb}=10^{-3}$ yr evolution
of all low-$M_c$ binaries shown in Figure \ref{fig:binary_evolve}c,d 
is typically dominated by torques set at 
$r_{\rm infl}$ located in the {\it gas pressure dominated} 
regime. Thus, GW phase signal will contain information only 
about the properties of the gas pressure dominated part of 
the disk, and will not inform us on the physics of the
inner, radiation pressure dominated regions.

We also note in this regard that the calculations of GW
shifts presented in Kocsis \etal (2011) \& Yunes \etal (2011)
should be revised to account for this non-locality of the 
disk torques on the binary.


\subsection{Spectra of disks around SMBH binaries.}  
\label{sect:EMprecursor}

\begin{figure}
\plotone{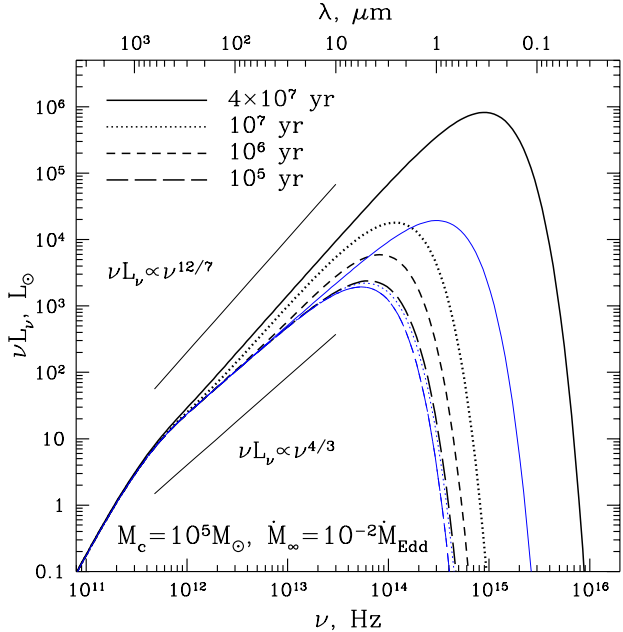}
\caption{
Evolution of the spectrum of a circumbinary disk evolving together 
with the central SMBH binary of mass $M_c=10^5$ M$_\odot$ and 
$q=1$, starting with initial semi-major axis $10^{-4}$ pc. Disk
has $\dot M_\infty=10^{-2}$ M$_{\rm Edd}$ far from the binary and 
extends out to $0.1$ pc. Its inner radius evolves together with 
the binary whose evolutionary track is shown in Figure 
\ref{fig:binary_evolve}d. Different curves show disk spectrum 
at different times labeled on the panel. Thick lines correspond to 
a disk self-consistently evolving under the action of the binary 
torque with no mass flow across the orbit of the secondary 
allowed ($\chi=0$). Thin (blue)
lines correspond to a constant $\dot M=\dot M_\infty$ disk occupying
the same region of space. See text for details.
\label{fig:spec_evolve}}
\end{figure}

Disks around SMBHs exhibit a set of 
observational signatures which distinguish them from 
the regular constant $\dot M$ accretion disks. We illustrate 
this difference in Figure \ref{fig:spec_evolve} by showing 
the spectral energy distribution (SED) of a circumbinary 
disk at different stages of the 
disk+binary evolution. This particular calculation 
assumes an equal mass, $M_c=10^5$ M$_\odot$ binary 
starting at $r_b(0)=10^{-4}$ pc,  surrounded
by a disk accreting gas at large separations at the rate 
$\dot M_\infty=10^{-2}\dot M_{\rm Edd}$; see Figure 
\ref{fig:binary_evolve}d for an evolutionary track of 
this system. 

One type of spectra shown by thick curves in Figure 
\ref{fig:spec_evolve} assumes radial distribution of 
the angular momentum flux in the disk $F_J(r,t)$ to be 
given by equation (\ref{eq:piece}) with $r_{\rm infl}(t)$ 
taken from the self-consistent calculations presented in 
Figure \ref{fig:binary_evolve}d. Another set of SEDs 
(thin curves) is computed for the same moments of 
time (and for the same values of $r_b(t)$) for a 
standard constant $\dot M=\dot M_\infty$ disk 
extending down to the binary orbit $r_b(t)$. In this case
$F_J(r,t)=\dot M_\infty\left(GM_c r\right)^{1/2}$ through
the entire disk. Both kinds of calculations assume an 
outer edge of the disk to lie at 
$0.1$ pc (in this calculation we disregard complications 
arising at large separations, which are mentioned in 
\S \ref{sect:global_properties}; the choice of the outer 
radius is not important to us). We now go over the details 
of these calculations.

After $t=10^5$ yr of evolution the binary semi-major axis 
has essentially not changed and the radius of its tidal 
influence has extended only out to 
$3\times 10^{-4}$ pc. Since this value of $r_{\rm infl}$
is close to $r_b$, binary torques affect only the very 
innermost part of the disk and the spectra computed
in two ways (thick and thin long-dashed curves in Figure 
\ref{fig:spec_evolve}) do not show significant difference.
At $\lambda=10-500$ $\mu$m both are well fit 
by a power law $\nu F_\nu\propto \nu^{4/3}$ typical for
a constant $\dot M$ disk.

At $t=10^6$ yr $r_b$ is still very close to $r_b(0)$, but
the effects of the binary torque have been viscously 
transmitted through the disk out to 
$r_{\rm infl}\approx 1.7\times 10^{-3}$ pc. This results in 
a factor of $\approx 4$ difference in the torque acting 
on the binary in two cases, and noticeably 
changes the spectrum of the disk in a self-consistent 
calculation: the peak wavelength of the spectrum shifts to 
a slightly shorter wavelength and the peak amplitude of 
$\nu F_\nu$ increases by a factor of 2 compared to $t=10^5$ 
yr. Note that the latter is a consequence only of the change
in the disk structure due to binary torques --- the inner disk
radius stays essentially the same in these two epochs. For the same 
reason there is not difference in the spectra computed assuming 
constant $\dot M$ disk (the thin curves for $t=10^5$ and 
$10^6$ essentially overlap in Figure \ref{fig:spec_evolve}).

At $t=10^7$ yr $r_b$ has shrunk to $8.7\times 10^{-5}$ pc,
while $r_{\rm infl}\approx 10^{-2}$ pc. Prior to this moment 
of time the disk in the immediate vicinity of the binary 
was in a gas pressure dominated state with free-free opacity,
and the radius of influence was also in the same regime.
At $t=10^7$ yr binary starts entering the gas pressure dominated 
part of the disk with $\kappa=\kappa_{es}$, see Figure 
\ref{fig:binary_evolve}d. While the
spectrum of a constant $\dot M$ disk is almost 
the same at this epoch, the SED of a self-consistently 
evolved disk (thick dotted curve) 
exhibits not only an increase in amplitude and a shift 
towards shorter wavelengths, but also a change in slope at 
$\lambda=10-500$
$\mu$m: in this range $\nu F_\nu$ is clearly steeper than
$\nu^{4/3}$. All that is again predominantly due to the 
evolution of the radial structure of the disk under the 
action of the binary torque.

Finally, at $t\approx 4\times 10^{7}$ yr binary orbit shrinks
to $10^{-5}$ pc, while $r_{\rm infl}\approx 0.03$ pc. Note that at this 
separation orbital evolution of the binary is dominated by
the GW emission rather than the disk torques, see Figure 
\ref{fig:binary_evolve}d. However, the binary
still causes the disk to evolve as long as its torque prevents 
the mass inflow into the inner cavity. 

Reduction
of $r_b$ leads to a change of a constant $\dot M$ disk SED ---
its peak is now around $1$ $\mu$m (thin solid line) and the 
peak of $\nu F_\nu$ is about an order of magnitude higher 
than before. But the variation of the spectrum of a 
self-consistently evolved disk (thick solid line) is far 
more dramatic --- it now peaks in the optical at $0.3$ $\mu$m 
and the peak 
value of $\nu F_\nu$ is $\approx 40$ times higher than for 
a constant $\dot M$ disk. Between $1$ $\mu$m and
$500$ $\mu$m the shape of the SED is well fit by 
$\nu F_\nu\propto \nu^{12/7}$, as expected for a constant
$\FJ$ disk (SC95), now occupying the inner 
third of the radial extent of the disk and 
accounting for most of its luminosity. 

To summarize, the SED of a disk affected by the torque of a central 
binary is steeper, brighter, and extends to shorter wavelengths 
than the SED of its constant $\dot M$ counterpart 
having the same inner radius and mass accretion rate 
$\dot M_\infty$ at large distances. 

These features (especially the steepness
of the spectrum) in principle make it possible to predict 
the existence of a compact central binary in a disk based 
on the broadband spectroscopy alone, even in the absence of 
other indications of a SMBH binary such as the double-peaked 
line profiles caused by the relative motion of the 
binary components. This method may be the only way of inferring
the presence of a binary in systems with face-on orientation. 
Relative brightness of disks affected by the binary torque
should facilitate the detection of such systems out to large 
distances.


\subsubsection{Sensitivity of the SED to mass inflow across
the secondary orbit.}
\label{sect:spec_inflow}

So far our calculations of the SMBH binary evolution
and electromagnetic signatures have explicitly assumed
that the binary torques completely prevent mass inflow
from the circumbinary disk across the orbit of the 
secondary. This allowed us to use the boundary condition
$\dot M(r_{in})=0$ in all our calculations, resulting in 
constant $F_J$ disk near the binary.

\begin{figure}
\plotone{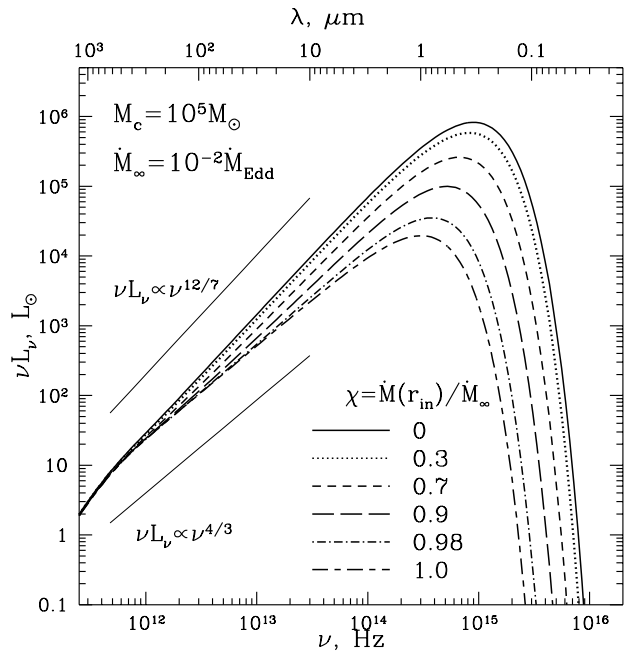}
\caption{
Dependence of the disk spectrum on the fraction of mass inflow 
$\chi=\dot M(r_{in})/\dot M_\infty$ that penetrates across the 
orbit of the secondary and leaves the circumbinary disk. 
Different curves correspond to different $\chi$ as labeled 
on the panel. Calculations are done for the situation depicted
by solid curves in Figure \ref{fig:spec_evolve}, i.e. after
$4\times 10^7$ yr of evolution of the disk+binary system shown
there. 
\label{fig:spec_compare}}
\end{figure}

If the zero inflow requirement is relaxed and 
$\dot M(r_{in})\neq 0$ then one can still construct both 
the steady state and the evolving self-similar solutions
as demonstrated in \S \ref{sect:steady} \& \ref{sect:self}.
Based on these solutions one can easily extend our results 
for the orbital evolution of the SMBH binary presented
in \S \ref{sect:inspiral} \& \ref{sect:inspiral_pars} to 
the case of non-zero mass inflow at the inner edge of the 
circumbinary disk.

Here we only show how the SED of the disk evolves
as one varies the transparency of the inner barrier presented
to the gas inflow by the binary torques. In Figure 
\ref{fig:spec_compare} we show disk spectra computed assuming 
different values of $\chi=\dot M(r_{in})/\dot M_\infty\le 1$. 
Instead of equation (\ref{eq:piece}) we now approximate the 
spatial distribution of $\FJ$ by the following simple formula:
\ba
F_J(r)= \left\{
\begin{array}{c}
\dot M_\infty l(r),~~~r>r_{\rm infl},\\
\dot M_\infty\left[l_{\rm infl}-\chi\left(l_{\rm infl}-
l(r)\right)\right],~~~r\le r_{\rm infl},
\end{array}
\right.
\label{eq:piece1}
\ea 
where $l(r)=\left(G M_c r\right)^{1/2}$, 
$l_{\rm infl}=\left(G M_c r_{\rm infl}\right)^{1/2}$. 
This prescription consists of two steady state solutions 
continuously matched at $r=r_{\rm infl}$ but with different 
$\dot M$ inside and outside of this point.
For $\chi=0$ this formula naturally reduces to equation 
(\ref{eq:piece}).

Our calculations assume a system for which a spectrum is
shown in Figure \ref{fig:spec_evolve}
at $t=4\times 10^7$ yr for $\dot M(r_{in})=0$, i.e. an equal mass, 
$M_c=10^5$ M$_\odot$ SMBH binary at $r_b=10^{-5}$ pc surrounded
by a disk with  
$\dot M_\infty=10^{-2}\dot M_{\rm Edd}$, and
$r_{\rm infl}\approx 0.03$ pc. We show the spectrum of only 
the {\it circumbinary} disk, i.e. in this calculation we do 
not account for the emission produced by an accretion disk(s) 
around the primary and/or secondary which should form when 
mass flows across the orbit of the secondary.

One can see that as the transparency of the tidal barrier 
$\chi$ increases towards unity the disk spectrum steadily 
approaches that of a constant $\dot M=\dot M_\infty$ disk.
This is not at all surprising because in the limit $\chi\to 1$
disk structure reduces to that of a constant $\dot M$
disk, see Figure \ref{fig:self-similar}. 
As a result the slope of the power law portion of the spectrum
steadily goes down from $12/7$ to
$4/3$ as $\chi$ is varied from $0$ to $1$. 

It is also clear from Figure \ref{fig:spec_compare} that SED 
is strongly affected compared to the case of a constant $\dot M$
($\chi=1$) disk even if only a small amount of inflowing mass gets 
accumulated at the inner edge of the disk. For example, peak 
amplitude of the disk spectrum for $\chi=0.98$ (implying that
only $2\%$ of the accreting mass gets stopped by the binary 
torques) is a factor of $2$ higher than in the $\chi=1$ case.
And if the tidal barrier allows penetration of only $30\%$
of the gas across the gap, the SED of the circumbinary disk 
is hardly distinguishable from that of a $\chi=0$ disk
with no gas inflow at $r_{in}$. 

These results imply that the broadband SED of the 
disk is a rather sensitive measure of even a small
amount of matter penetrating into the cavity cleared by the 
SMBH binary. Coupled with the measurements of the SED of the
accretion disk(s), which may form around each of the binary 
components if $\chi\neq 0$, these observations can inform
us on the efficiency of the binary torques at clearing 
a clean cavity at the center of the system.  


\section{Summary.}  
\label{sect:summ}

In this work we explored the coupled evolution of a SMBH binary 
and a gaseous disk around it. Disk properties 
(surface density, temperature, etc.) evolve under the action
of binary torques, which constrain the flow in the inner part 
of the disk. To study this problem we have re-formulated evolution
equations in terms of the angular momentum flux $\FJ$. This
significantly simplifies treatment of the steady state disk
structure, when $\FJ$ is a simple linear function 
of the specific angular momentum $l$. 

We derived the disk properties as a function of $\FJ$
in different physical regimes that may be realized in circumbinary
disks around SMBH binaries. We demonstrated that radiation 
pressure can limit the value of $\FJ$ in disks around massive
SMBH binaries by making the disk geometrically thick and 
susceptible to launching a radiation-driven wind. 

When the external mass supply to the disk at large 
distances is not matched at the inner edge of the disk because of 
the binary torques, the disk evolves towards establishing a 
quasi-steady state in the inner region, where the local 
viscous timescale is shorter than the evolution time of the 
system.
Viscous angular momentum flux in the inner disk steadily
grows in time, which accelerates 
orbital evolution of the binary.

We explored the dependence of this general picture  
on the system parameters (mass of the binary, mass accretion rate
through the disk, etc.) and found the following in agreement 
with previous studies. 
\begin{enumerate}
\item Tidal coupling to a circumbinary disk can substantially
(by orders of magnitude) shorten the lifetime of the binary
(IPP; Lodato \etal 2009; Haiman \etal 2009). 
\item For a long period before the GW emission takes over, the binary
evolves in the limit when the mass of the secondary is much larger 
than the local disk mass (Haiman \etal 2009).
\item Disk-driven evolution of the binary can be measurable by the 
space based gravitational wave antennae for low $q$
systems with relatively low $M_c$ (Kocsis etal 2011; Yunes \etal 2011).
\item Spectrum of the disk affected by the binary torques is
different from that of a conventional constant $\dot M$ disk: it 
extends to shorter wavelengths and more power is emitted. Instead
of $\nu F_\nu\propto \nu^{4/3}$ the SED of a circumbinary disk 
exhibits a power law segment with $\nu F_\nu\propto \nu^{12/7}$
(SC95). 
\end{enumerate} 

We also obtain a number of new results, summarized below.
\begin{enumerate}
\item Self-consistent evolution of the disk resulting in a pile-up
of mass at its inner edge accelerates the orbital evolution of the
binary.
\item Disk-binary coupling has a non-local character: the torque
acting on the binary is determined by the state of the disk far from
the binary, at the radius of influence $r_{\rm infl}$, which steadily 
increases in time, rather than by the disk properties in the 
immediate vicinity of the binary. 
\item Evolution of the binary orbits exhibits a phenomenon of 
hysteresis --- dependence of the evolution on the past history 
of the system, which is caused by the non-locality of the 
disk-binary coupling.
\item Radiation pressure can strongly affect the
disk structure even in cases when the mass accretion rate at large 
distances (in the constant $\dot M$ portion of the disk) is considerably
sub-Eddington.
\item Gas overflow across the orbit of the secondary 
affects binary mainly (or only) during the GW-dominated
phase of its orbital evolution and is most important for 
low $q$, high $\dot M_\infty$ systems.
\item Spectra of circumbinary disks strongly depend on the 
ability of accreting gas to cross the orbit of the secondary or
otherwise leave the system, thus giving rise to a non-zero value 
of $\dot M$ at the inner edge of the disk.
\end{enumerate}

This list clearly implies that properly accounting for the fully
self-consistent, time-dependent evolution of circumbinary disks
is crucial for understanding gas-assisted SMBH mergers. This 
general conclusion will hopefully inspire re-evaluation of 
some of the existing results for the orbital evolution of SMBH 
binaries and their observational manifestations, both in the 
electromagnetic and the GW domains. 
Results of this work can also be extended to studying circumbinary 
disks around stellar mass binaries.


\acknowledgements 

I am indebted to Pavel Ivanov, Bence Kocsis and Zoltan Haiman
for careful reading of the manuscript, open exchange of 
opinions, and a number of useful suggestions. 
The financial support for this work is provided by 
the Sloan Foundation, NASA grant NNX08AH87G, and NSF grant 
AST-0908269.



\appendix

\section{Scaling relations for arbitrary power law opacity.}  
\label{sect:gen_opacity}

Here we summarize scaling relations for disk properties that 
result when opacity is a power law function of gas temperature 
$T$ and density $\rho$ (here taken to be represented by their 
midplane values):
\ba
\kappa=\kappa_0\rho^{\mu_1}T^{\mu_2}.
\label{eq:op_law}
\ea
We make two additional assumptions regarding disk properties:
(1) disk is optically thick and (2) radiation pressure is 
negligible compared to the gas pressure (radiation pressure 
dominated case is described by equations 
(\ref{eq:h_rad_es})-(\ref{eq:T_b=1})).

Combining equations (\ref{eq:Fnu}), (\ref{eq:T_gen}), 
(\ref{eq:rad_transfer}), \& (\ref{eq:surfdens}), and 
$\nu=\alpha c_s^2/\Omega$ valid in the gas pressure 
dominated regime one finds 
\ba
\Sigma(r)&=&\left[\frac{2^{2(4+\mu_1)}}{3^{10+\mu_1-2\mu_2}
\pi^{6+\mu_1-2\mu_2}}\left(\frac{\sigma}{\kappa_0}\right)^2
\left(\frac{\mu}{k}\right)^{2(4-\mu_2)}
\frac{\FJ^{6+\mu_1-2\mu_2}}{\alpha^{8+\mu_1-2\mu_2}
(GM_c)^{1+\mu_1}}\right]^{1/\epsilon}r^{-(9-\mu_1-4\mu_2)/\epsilon},
\label{eq:sig_general}
\\
T(r)&=&\left[\frac{2^{-(4+\mu_1)}}{3^{\mu_1}
\pi^{2+\mu_1}}\frac{\kappa_0}{\sigma}
\left(\frac{\mu}{k}\right)^{(2+3\mu_1)/2}
\frac{\FJ^{2+\mu_1}(GM_c)^{(1+\mu_1)/2}}{\alpha^{1+\mu_1}}
\right]^{2/\epsilon}r^{-(11+7\mu_1)/\epsilon},
\label{eq:T_general}
\\
\frac{h(r)}{r}&=&\left[\frac{2^{-(4+\mu_1)}}{3^{\mu_1}
\pi^{2+\mu_1}}\frac{\kappa_0}{\sigma}
\left(\frac{k}{\mu}\right)^{4-\mu_2}
\frac{\FJ^{2+\mu_1}}{\alpha^{1+\mu_1}(GM_c)^{(9+2\mu_1-2\mu_2)/2}}
\right]^{1/\epsilon}r^{-(1+4\mu_1+2\mu_2)/(2\epsilon)},
\label{eq:hr_general}
\ea
where $\epsilon\equiv 10+3\mu_1-2\mu_2$.

Using these results we can derive an expression for the
diffusion coefficient $D_J$ in the form (\ref{eq:PL_diff}):
\ba
D_{J,0}&=&\frac{3}{4}\left[\frac{2^{-4-\mu_1}}
{3^{\mu_1}\pi^{2+\mu_1}}\alpha^{(8+\mu_1-2\mu_2)/2}
\frac{\kappa_0}{\sigma}\left(\frac{k}{\mu}\right)^{4-\mu_2}
(GM_c)^{6+4\mu_1}\right]^{2/(10+3\mu_1-2\mu_2)},
\label{eq:DJ0_PLgen}
\\
d&=& \frac{2(2+\mu_1)}{10+3\mu_1-2\mu_2},~~~~~
p=-\frac{12+11\mu_1+2\mu_2}{10+3\mu_1-2\mu_2}.
\label{eq:indices_PLgen}
\ea
See Lyubarskij \& Shakura (1987) for similar results.


\section{Summary of the diffusion coefficient behavior
in different regimes.}  
\label{sect:D_J}

Here we summarize the behavior of the diffusion coefficient 
$D_J$ in the power law form  (\ref{eq:PL_diff}) and of the
self-similar exponent $n$ defined by equation 
(\ref{eq:self_sim_form}) for different 
objects and in different regimes explored in this work (see 
also Lyubarskij \& Shakura 1987). 

In the radiation pressure dominated case (\S \ref{sect:rad}) 
one finds for $b=0$
\ba
D_{J,0}=\frac{3}{2^4\pi^2}\alpha\frac{\kappa_{es}^2(GM_c)^4}{c^2},
~~~~~d= 2,~~~~~
p=-7,~~~~~n=\frac{1}{7}.
\label{eq:DJ_b=0}
\ea
and for $b=1$
\ba
D_{J,0}=\frac{3}{4}\left[\frac{1}{2^{4}\pi^2}
\left(\frac{k}{\mu}\right)^4
\frac{(GM_c)^6\kappa_{es}\alpha^4}{\sigma}
\right]^{1/5},~~~~~d=\frac{2}{5},~~~~~
p=-\frac{6}{5},~~~~~n=\frac{5}{14}.
\label{eq:DJ_b=1}
\ea
Expression (\ref{eq:DJ_b=1}) also holds true for
the gas pressure dominated case with $\kappa=\kappa_{es}$ 
(\S \ref{sect:gas_es}).

In the gas pressure dominated case with $\kappa=\kappa_{ff}$ 
(\S \ref{sect:gas_ff})
\ba
D_{J,0}=\frac{3}{4}(GM_c)\left[\frac{2^{-5}}{3\pi^{3}}
\frac{\kappa_0}{\sigma}\left(\frac{k}{\mu}\right)^{15/2}
\alpha^8\right]^{1/10},~~~~~d=\frac{3}{10},~~~~~
p=-\frac{4}{5},~~~~~n=\frac{2}{5}.
\label{eq:DJ_ff}
\ea

\end{document}